\documentclass[a4paper,fleqn,usenatbib]{mnras}

\usepackage[T1]{fontenc}
\usepackage{ae,aecompl}

\usepackage{graphicx}	
\usepackage{amsmath}	
\usepackage{amssymb}	
\usepackage[dvipsnames]{xcolor}

\newcommand{\hi}{H~\textsc{i}}
\newcommand{\hii}{H~\textsc{ii}}
\newcommand{\hei}{He~\textsc{i}}
\newcommand{\heii}{He~\textsc{ii}}
\newcommand{\lya}{Lyman-$\alpha$}

\title[The Sherwood-Relics simulations]
{The Sherwood-Relics simulations: overview and impact of patchy reionization and pressure smoothing on the intergalactic medium}

\author[Puchwein et al.]{Ewald Puchwein$^{1}$\thanks{E-mail: epuchwein@aip.de},
James S. Bolton$^{2}$,
Laura C. Keating$^{1}$,
Margherita Molaro$^{2}$, \newauthor
Prakash Gaikwad$^{3}$,
Girish Kulkarni$^{4}$,
Martin G. Haehnelt$^{5}$, 
Vid Ir\v{s}i\v{c}$^{5}$, \newauthor
Tom\'a\v{s} \v{S}oltinsk\'y$^{2}$,
Matteo Viel$^{6,7,8,9}$,
Dominique Aubert$^{10}$, 
George D. Becker$^{11}$ \newauthor
and Avery Meiksin$^{12}$
\\
$^{1}$Leibniz-Institut f\"ur Astrophysik Potsdam, An der Sternwarte 16, 14482 Potsdam, Germany\\
$^{2}$School of Physics and Astronomy, University of Nottingham, University Park, Nottingham, NG7 2RD, UK\\
$^{3}$Max-Planck-Institut für Astronomie, Königstuhl 17, D-69117 Heidelberg, Germany\\
$^{4}$Tata Institute of Fundamental Research, Homi Bhabha Road, Mumbai 400005, India\\
$^{5}$Kavli Institute for Cosmology and Institute of Astronomy, Madingley Road, Cambridge, CB3 0HA, UK\\
$^{6}$SISSA - International School for Advanced Studies, Via Bonomea 265, I-34136 Trieste, Italy\\
$^{7}$IFPU, Institute for Fundamental Physics of the Universe, Via Beirut 2, I-34151 Trieste, Italy\\
$^{8}$INFN, Sezione di Trieste, Via Valerio 2, I-34127 Trieste, Italy\\
$^{9}$INAF - Osservatorio Astronomico di Trieste, Via G. B. Tiepolo 11, I-34143 Trieste, Italy\\
$^{10}$Observatoire Astronomique de Strasbourg, 11 rue de l’Universite, 67000 Strasbourg, France\\
$^{11}$Department of Physics and Astronomy, University of California, Riverside, CA, 92521, USA\\
$^{12}$Institute for Astronomy, University of Edinburgh, Blackford Hill, Edinburgh, EH9 3HJ, UK\\
}

\date{Accepted XXX. Received YYY; in original form ZZZ}

\pubyear{2022}

\begin{document}
\label{firstpage}
\pagerange{\pageref{firstpage}--\pageref{lastpage}}
\maketitle

\begin{abstract}
 We present the Sherwood-Relics simulations, a new suite of large cosmological hydrodynamical simulations aimed at modelling the intergalactic medium (IGM) during and after the cosmic reionization of hydrogen. The suite consists of over 200 simulations that cover a wide range of astrophysical and cosmological parameters. It also includes simulations that use a new lightweight hybrid scheme for treating radiative transfer effects. This scheme follows the spatial variations in the ionizing radiation field, as well as the associated fluctuations in IGM temperature and pressure smoothing. It is computationally much cheaper than full radiation hydrodynamics simulations and circumvents the difficult task of calibrating a galaxy formation model to observational constraints on cosmic reionization. Using this hybrid technique, we study the spatial fluctuations in IGM properties that are seeded by patchy cosmic reionization. We investigate the relevant physical processes and assess their impact on the $z > 4$ \lya\ forest. Our main findings are: (i) Consistent with previous studies patchy reionization causes large scale temperature fluctuations that persist well after the end of reionization, (ii) these increase the \lya\ forest flux power spectrum on large scales, and (iii) result in a spatially varying pressure smoothing that correlates well with the local reionization redshift. (iv) Structures evaporated or puffed up by photoheating cause notable features in the \lya\ forest, such as flat-bottom or double-dip absorption profiles. 
\end{abstract}




\section{Introduction}

The ionizing UV emission produced by the first stars and galaxies in the high-redshift Universe transforms the surrounding IGM from a neutral gas to a highly ionized plasma. At the same time, it is photoheated from a few Kelvin to $\sim 10^4$ K. As this process of cosmic reionization proceeds, ionized regions grow, start to overlap, and eventually become volume filling \citep[e.g., see reviews by ][]{rauch1998,meiksin2009,mcquinn2016}. This inherently inhomogeneous process results in an almost fully ionized IGM. Neutral gas is only present in dense regions that can self-shield from the cosmic UV background.

The photoheating provided by cosmic reionization increases the gas pressure in the IGM such that it becomes dynamically relevant on small scales. While gravity still dominates the formation of structures in the baryonic density field on large scales, small scales in the IGM are notably affected by the hydrodynamic reaction to the photoheating. Overpressurized regions expand, thereby erasing structure on small scales \citep[e.g.,][]{1998MNRAS.296...44G,2000MNRAS.315..600T,KulkarniEtAl2015,rorai2017,WuEtAl2019,KatzEtAl2020,2021ApJ...923..161N}. How exactly this process proceeds is sensitive to the amount of heating provided by reionization, but also to the initial properties of the neutral gas, such as the relative streaming velocity between baryons and dark matter and the amount of preheating by X-rays penetrating into neutral regions \citep[see, e.g.,][]{2018MNRAS.474.2173H,2021ApJ...908...96P,2022MNRAS.513..117L}. Understanding pressure smoothing is relevant both for explaining the formation (or lack of formation) of galaxies in low mass halos, as well as the properties of the IGM during and after cosmic reionization. Here, we concentrate on the latter, i.e. on studying the immediate impact of patchy cosmic reionization on the IGM, as well as the \textit{relic} signatures of patchy reionization that persist for a significant amount of time in the post-reionization IGM \citep[e.g.,][]{LidzMalloy2014,DAloisioEtAl2015,KeatingEtAl2018,OnorbeEtAl2019,WuEtAl2019,2019MNRAS.487.1047M,MonteroCamacho2020,MolaroEtAl2022}.

The IGM is most readily observed in absorption, in particular using the \lya\ line of neutral hydrogen, which imprints a forest of absorption lines on the spectra of background quasars. The structure of this \lya\ forest on small scales is affected by the instantaneous temperature of the IGM via the Doppler broadening of the lines, by its thermal history via the pressure smoothing, and by the small scale structure in the dark matter density field via gravitational interaction. The latter has been exploited by using the \lya\ forest on small scales to probe the free streaming scale of dark matter particles \citep[e.g.,][]{2017PhRvD..96b3522I,2021PhRvL.126g1302R}, which for thermal relic dark matter is directly related to the dark matter particle mass. Such \lya\ forest constraints on dark matter are best derived at high-redshift when the relevant scales are not yet completely dominated by mode coupling due to non-linear structure growth and when the \lya\ forest is sensitive to low (less non-linear) densities. Furthermore, for fixed comoving free-streaming length the cut-off in velocity space is at larger scales/smaller $k$ at higher redshift, and thus - at least in principle - easier to detect. At very high redshift, when entering the epoch of reionization, the \lya\ forest becomes completely opaque. Hence, the sweet spot for dark matter constraints is in the redshift range $4 \lesssim z \lesssim 5.5$, or in other words shortly after reionization. Understanding the thermal state and pressure smoothing of the IGM in this epoch is, hence, also important for probing dark matter.

Interpreting observations of the IGM typically relies heavily on comparison to cosmological hydrodynamical simulations. The main differences between the various simulations used for this purpose are how the ionizing sources, i.e. the galaxies, and the ionizing radiation fields are treated. In the most simple approach, these are not treated explicitly \citep[see, e.g.,][for recent works]{Sherwood,Rossi_2020,Chabanier_2020,2021JCAP...04..059W,Villasenor_2021}. Instead an external model for the ionizing UV background (UVB) is used, typically a spatially homogeneous, time varying UVB model. Such models are obtained by integrating the ionizing emission of stars and active galactic nuclei based on empirical constraints of their abundance and of the opacity of the IGM \citep[e.g.,][]{2009ApJ...703.1416F,HaardtMadau2012,onorbe2017,PuchweinEtAl2019,khaire2019,2020MNRAS.493.1614F}. When focusing on the low-density IGM probed by the \lya\ forest, it is possible to neglect other forms of feedback from galaxies, as it does typically not reach the relevant low-density regions of the IGM at $z\gtrsim 4$ \citep[although feedback will start to play a role by $z\sim 2$, see e.g.][]{Theuns_2002,Viel_2013,Chabanier_2020}. Hence, when using an external UVB model, one can reasonably ignore galaxy formation altogether in cosmological hydrodynamical simulations of the low-density IGM \citep{QuickLya}. Despite their simplicity and low computational cost, such simulations are also in remarkably good agreement with the observed properties of the \lya\ forest at $z\lesssim4$, i.e. well after the end of reionization  \citep[e.g.][]{Sherwood}.

During patchy cosmic reionization, the ionizing radiation field is, however, highly inhomogeneous and simulations with a spatially homogeneous UVB model fail to reproduce the observed properties of the \lya\ forest, such as the fluctuations of its opacity on large scales \citep[e.g.][]{BeckerEtAl2015,bosman2018,eilers2018,Zhu_2021,bosman2022}. To overcome these problems, the spatial distribution of ionizing sources and the resulting spatial fluctuations in the UV radiation field need to be modelled. This can either be done on the fly in full galaxy formation simulations with radiative transfer coupled to the hydrodynamics, or by doing the radiative transfer in post-processing, typically using a simpler model of the ionizing source populations.

The latter approach is computationally much cheaper and can avoid all the complications of realistically modelling the galaxy population and the escape of ionizing radiation. Empirically constrained source models can be used to ``paint'' ionizing sources on the simulated dark matter halos. A weakness of this approach is that it may miss many details of the source population, such as the bursty nature of ionizing radiation production and escape. Nevertheless, the source luminosities can be calibrated such that the amount of ionizing radiation reaching the IGM is adequate for bringing the reionization history and the simulated properties of the \lya\ forest in agreement with observational constraints. Calibrating in this manner makes a detailed modelling of the ionizing radiation escape from high-density regions in the simulation unnecessary. Hence, the post-processing radiative transfer can be done at coarser resolution using an uniformly spaced grid. This also allows efficient parallelization on GPUs \citep[e.g.,][]{Aton2010}, making this approach numerically cheap. Such calculations are very successful in matching the properties of the \lya\ forest on large scales during and directly after cosmic reionization \citep[e.g.,][]{KulkarniEtAl2019}.       

On the downside, post-processing radiative transfer neglects the hydrodynamic reaction of the IGM to the inhomogeneous photoheating. The thermal and ionization states are only re-calculated in post processing so that the heating is not coupled to the hydrodynamics. Hence, a self-consistent modelling of pressure smoothing is not possible with this approach. Furthermore, the thermal and ionization states are typically stored on a static grid so that the thermal energy injected by photoheating as well as the ionization state are not advected with the gas flow. Finally, post-processing radiative transfer codes that follow only heating by the UV radiation field miss other heating mechanisms that are present in a hydrodynamic simulation, such as shock heating in and around forming structures. All of these issues can be fixed by doing the radiative transfer in a fully coupled manner on the fly. Such radiation-hydrodynamics simulations can be done with external ionizing sources to study the reaction of the IGM \citep[e.g.,][]{2016ApJ...831...86P,DAloisioEtAl2020} or following the formation of high-redshift galaxy populations and the escape of ionizing radiation from them self-consistently. The latter is, however, computationally very expensive. Usually some corners need to be cut to make this feasible at all, e.g., doing the radiative transfer with a reduced speed of light. In addition, the modelling of galaxy formation in such simulations is highly uncertain, in particular during the epoch of reionization where only very limited observational constraints on the galaxy population are available, making sanity checks on the simulated population difficult. This is further exacerbated by the fact that the escape of ionizing radiation depends on the detailed properties of the interstellar medium, which is very challenging to model faithfully in cosmological simulations. Despite these difficulties, there has been major progress in this direction in the last years. For example, the CROC \citep{2014ApJ...793...29G}, Sphinx \citep{Sphinx}, Technicolor Dawn \citep{Finlator2018}, CoDa \citep{ocvirk2016,ocvirk2020,lewis2022} and Thesan \citep{Thesan,garaldi2022,2022MNRAS.512.3243S} collaborations were able to perform fully-coupled radiation-hydrodynamics simulations of cosmological volumes along with a modelling of galaxy formation.

In this work, we focus on the low-density IGM probed by the \lya\ forest. We build on our previous Sherwood simulation project \citep{Sherwood}, which showed excellent agreement with various statistics of the \lya\ forest at $2 \lesssim z \lesssim 5$. Our aim is to produce a simulation suite that samples a wide range of astrophysical and cosmological parameters and that can be used for parameter inference when compared to the observed \lya\ forest. At the same time, we aim to overcome limitations
at the high redshift end where the \textit{relic} signatures that a recently completed patchy cosmic reionization has imprinted on the IGM become increasingly important. To this end, we introduce a new hybrid radiative transfer/hydrodynamical simulation technique that aims at combining many of the positive aspects of post-processing radiative transfer and fully-coupled radiation-hydrodynamics simulations while avoiding some of their major downsides. 

In Sec.~\ref{sec:methods} of this manuscript, we will describe the simulation methods in detail. Sec.~\ref{sec:grid_pow_spec} will give an overview of how varying different parameters affects the \lya\ forest on different scales. Sec.~\ref{sec:patchy_results} focuses on our hybrid patchy reionization simulations, with the thermal state of the IGM being discussed in \ref{sec:thermal_properties}, the modulation of the \lya\ forest on large scales in \ref{sec:modulation_lya},  the spatially varying pressure smoothing of the IGM in \ref{sec:pressure_smooth}, and its direct imprints on the \lya\ forest in \ref{sec:lya_lines_rings}. We summarize our results in Sec.~\ref{sec:summary}.

In the Sherwood-Relics project, we assume for our baseline model the same $\Lambda$CDM cosmology as in the Sherwood project. We, thus, use $\Omega_\textrm{m} \! = \! 0.308$, $\Omega_\Lambda \! = \! 0.692$, $\Omega_\textrm{b} \! = \! 0.0482$, $h \! = \! 0.678$, $\sigma_8 \! = \! 0.829$, and $n \! = \! 0.961$ unless specifically mentioned otherwise.

\section{Methods}
\label{sec:methods}

\subsection{The Sherwood-Relics simulation suite}
\label{sec:relics_suite}

\begin{table*}
  \centering
   \caption{Summary of the Sherwood-Relics simulation suite. The columns contain
     (i) the box size, (ii) the cube root of the (initial) gas particle number, 
     (iii) the dark matter and (iv) gas mass resolution, (v) the gravitational softening, as well
     as several properties that have been varied between the different runs.
     This includes (vi) the dark matter model (cold dark matter or warm dark matter particle mass),
     (vii) the \hi\ photoheating normalization factor
     (see Sec.~\ref{sec:sims_homo_uvb}), (viii) the global reionization redshift $z_\textrm{r}$ at which \hi\ reionization
     completes (defined as the redshift when the volume-averaged neutral fraction in the simulation falls below $10^{-3}$), and (ix) the redshift $z_{\rm mid}$ when the volume-averaged neutral fraction reaches $0.5$. The last column provides (x) the number of runs $N_\textrm{runs}$, as well as any further comments.}
  \resizebox{\textwidth}{!}{%
  \begin{tabular}{|c|c|c|c|c|c|c|c|c|c|c}
    \hline
    \hline
    Box size & $N^{1/3}$ & $M_{\rm dm}$ & $M_{\rm gas}$ & $l_{\rm soft}$ & Dark matter & HI photoheating & $z_\textrm{r}$ & $z_\textrm{mid}$ &  $N_{\rm runs}$ (comments) \\
    $[h^{-1} {\rm cMpc}]$ & & $\rm [h^{-1}\,M_{\odot}]$ & $\rm [h^{-1}\,M_{\odot}]$ & $\rm\,[h^{-1}\,ckpc]$ & & factor & & & & \\
    \hline
    \multicolumn{9}{c}{\textbf{Simulations with a homogeneous UV background}}\\
    \hline
     40 & 2048 & $5.37\times 10^{5}$ & $9.97\times 10^{4}$ & 0.78 & CDM, 2 keV, 3 keV, 4 keV & 0.5, 1, 2 & 6.0 & 7.7 & 12\\ 
     40 & 2048 & $5.37\times 10^{5}$ & $9.97\times 10^{4}$ & 0.78 & CDM & 1 & 5.4, 6.7, 7.4 & 7.1, 8.4, 9.1 & 3 \\
     40 & 2048 & $5.37\times 10^{5}$ & $9.97\times 10^{4}$ & 0.78 & CDM & tailored heating & 5.3, 5.7, 6.0, 6.6 & 7.2, 7.5, 7.3, 8.0 & 4 (matched to patchy)\\
       40 & 2048 & $5.37\times 10^{5}$ & $9.97\times 10^{4}$ & 0.78 & CDM & no photoheating & no ionization & no ionization  & 1 (adiabatic)\\

     80 & 2048 & $4.30\times 10^{6}$ & $7.97\times 10^{5}$ & 1.56 & CDM & 1 & 6.0 & 7.7 & 1\\
     160 & 2048 & $3.44\times 10^{7}$ & $6.38\times 10^{6}$ & 3.13 & CDM & tailored heating & 5.3 & 7.2 & 1 \\
     \hline
       5 & 1280 & $4.30\times 10^{3}$ & $7.97\times 10^{2}$  & 0.17  & CDM & 1 & 6.0 & 7.7  & 1\\
     10 & 1280 & $3.44\times 10^{4}$ & $6.38\times 10^{3} $ & 0.31 & CDM & 1 & 6.0 & 7.7 & 1\\
      20 & 1280 & $2.75\times 10^{5}$ & $5.10 \times 10^{4}$  & 0.63 & CDM & 1 & 6.0 & 7.7  & 1\\
    \hline
      5 & 1024 & $8.39\times 10^{3}$ & $1.56\times 10^{3}$  & 0.20  & CDM & 1 & 6.0 & 7.7 & 1\\
     10 & 1024 & $6.72\times 10^{4}$ & $1.25\times 10^{4}$ & 0.39 & CDM, 2 keV, 3 keV, 4 keV & 0.5, 1, 2 & 5.4, 6.0, 6.7, 7.4  & 7.1, 7.7, 8.4, 9.1 & 25 \\
     20 & 1024 & $5.37\times 10^{5}$ & $9.97\times 10^{4}$ & 0.78 & CDM, 2 keV, 3 keV, 4 keV & 0.5, 1, 2 & 5.4, 6.0, 6.7, 7.4 & 7.1, 7.7, 8.4, 9.1 & 48 \\
     20 & 1024 & $5.37\times 10^{5}$ & $9.97\times 10^{4}$ & 0.78 & 8 keV, 12 keV & 0.5, 1, 2 & 5.4, 6.0, 6.7, 7.4 & 7.1, 7.7, 8.4, 9.1 & 24 \\
     20 & 1024 & $5.37\times 10^{5}$ & $9.97\times 10^{4}$ & 0.78 & CDM & different $\gamma$ & 6.0 & 7.7 & 3 \\
     20 & 1024 & $5.37\times 10^{5}$ & $9.97\times 10^{4}$ & 0.78 & 1 keV & 1 & 6.0 & 7.7 & 1 \\
     20 & 1024 & $5.37\times 10^{5}$ & $9.97\times 10^{4}$ & 0.78 & CDM & 0.5, 1, 2 & 5.4, 6.0, 6.7, 7.4 & 7.1, 7.7, 8.4, 9.1 & 24 (different $\sigma_8$)\\
     20 & 1024 & $5.37\times 10^{5}$ & $9.97\times 10^{4}$ & 0.78 & CDM & 0.5, 1, 2 & 5.4, 6.0, 6.7, 7.4 & 7.1, 7.7, 8.4, 9.1 & 24 (different $n_\mathrm{s}$) \\
     40 & 1024 & $4.30\times 10^{6}$ & $7.97\times 10^{5}$ & 1.56 & CDM, 2 keV, 3 keV, 4 keV & 0.5, 1, 2 & 5.4, 6.0, 6.7, 7.4 & 7.1, 7.7, 8.4, 9.1 & 48 \\
     40 & 1024 & $4.30\times 10^{6}$ & $7.97\times 10^{5}$ & 1.56 & CDM & different $\gamma$ & 6.0 & 7.7 & 3 \\
      40 & 1024 & $4.30\times 10^{6}$ & $7.97\times 10^{5}$ & 1.56 & 1 keV & 1 & 6.0 & 7.7 & 1 \\
     40 & 1024 & $4.30\times 10^{6}$ & $7.97\times 10^{5}$ & 1.56 & CDM & 0.5, 1, 2 & 6.0 & 7.7 & 12 (different $\sigma_8$)\\
     40 & 1024 & $4.30\times 10^{6}$ & $7.97\times 10^{5}$ & 1.56 & CDM & 0.5, 1, 2 & 6.0 & 7.7 & 12 (different $n_\mathrm{s}$) \\
     80 & 1024 & $3.44\times 10^{7}$ & $6.38\times 10^{6}$ & 3.13 & CDM & 1 & 6.0 & 7.7 & 1\\
    \hline
    5 & 768 & $1.99\times 10^{4}$ & $3.69\times 10^{3}$  & 0.26  & CDM & 1 & 6.0 & 7.7 & 1\\
     10 & 768 & $1.59\times 10^{5}$ & $2.95\times 10^{4}$ & 0.52 & CDM & 1 & 6.0 & 7.7 & 1 \\
     20 & 768 & $1.27\times 10^{6}$ & $2.36\times 10^{5}$ & 1.04 & CDM & 1 & 6.0 & 7.7 & 1 \\
      40 & 768 & $1.02\times 10^{7}$ & $1.89\times 10^{6}$ & 2.08 & CDM & 1 & 6.0 & 7.7 & 1 \\
       80 & 768 & $8.15\times 10^{7}$ & $1.51\times 10^{7}$ & 4.17 & CDM & 1 & 6.0 &  7.7 & 1 \\
    \hline
    5 & 512 & $6.72\times 10^{4}$ & $1.25\times 10^{4}$  & 0.39  & CDM & 1 & 6.0 & 7.7 & 1 \\
     10 & 512 & $5.37\times 10^{5}$ & $9.97\times 10^{4}$ & 0.78 & CDM, 2 keV, 3 keV, 4 keV  & 0.5, 1, 2 & 5.4, 6.0, 6.7, 7.4 & 7.1, 7.7, 8.4, 9.1 &  25 \\
     20 & 512 & $4.30\times 10^{6}$ & $7.97\times 10^{5}$ & 1.56 & CDM & 1 & 6.0 & 7.7 &  1 \\
     40 & 512 & $3.44\times 10^{7}$ & $6.38\times 10^{6}$ & 3.13 & CDM & 1 & 6.0 & 7.7 & 1 \\
       80 & 512 & $2.78\times 10^{8}$ & $5.10\times 10^{7}$ & 6.25 & CDM & 1 & 6.0  & 7.7 & 1 \\
    \hline
    \multicolumn{9}{c}{\textbf{Simulations with an inhomogeneous radiation field (patchy reionization)}}\\
    \hline
     40 & 2048 & $5.37\times 10^{5}$ & $9.97\times 10^{4}$ & 0.78 & CDM & - & 5.3, 5.7,  6.0, 6.6 & 7.2, 7.5, 7.3, 8.0 & 4 \\
    160 & 2048 & $3.44\times 10^{7}$ &  $6.38\times 10^{6}$ &  3.13 & CDM & - & 5.3 & 7.2 & 1 \\
    \hline
    \hline
  \end{tabular}}
  \label{tab:sims}
\end{table*}

The Sherwood-Relics suite that we present here builds upon the Sherwood simulation project \citep{Sherwood}. In particular,
we simulate the same volumes and compute mock \lya\ forest absorption spectra on-the-fly in the same way. 
We, however, expand the sampled space of astrophysical and cosmological parameters significantly. We explore different reionization and heating histories, different values for the most relevant cosmological parameters for the \lya\ forest ($\sigma_8$ and $n_\textrm{s}$), and investigate the impact of the patchiness of cosmic reionization on the high-redshift IGM ($z \gtrsim 4$). Furthermore, since the sweet spot for \lya\ forest constraints on dark matter also falls in this redshift range, we simulate models with a range of different dark matter free streaming scales. Table~\ref{tab:sims} provides an overview of the over 200 different simulations performed for this project. 

On a technical level, the main improvements compared to the Sherwood simulations are a non-equilibrium thermo-chemistry solver and an improved treatment of the ionizing radiation fields. The latter includes both simulations with an improved time-dependent but spatially-homogeneous UV background model (as detailed in Sec.~\ref{sec:sims_homo_uvb}, based on \citealt{PuchweinEtAl2019}), as well as with a new hybrid post-processing radiative transfer/hydrodynamical simulation treatment of patchy reionization (as introduced in Sec.~\ref{sec:sims_hybrid_method}). 

Several studies have already made use of the Sherwood-Relics suite. Using our hybrid simulations, \citet{MolaroEtAl2022} have derived corrections for the impact of the patchiness of reionization on the \lya\ forest flux power spectrum. These corrections can be applied to conventional simulations with homogeneous UVB models and will be used together with our large grid of homogeneous UVB simulations in forthcoming studies. Other aspects of our hybrid simulations have also already been explored in \citet{2020MNRAS.494.5091G}, focusing on the properties of \lya\ transmission spikes, and in \citet{2021MNRAS.506.5818S}, predicting the 21-cm forest. \citet{lamberts2022} have used our baseline homogeneous UVB simulation as a reference for the expected thermal history during \heii\ reionization. The main science focus of this work is to provide a comprehensive introduction to the Sherwood-Relics simulations and to investigate the physical processes by which patchy reionization affects the IGM and \lya\ forest.

\subsection{The simulation code}
\label{sec:sim_code}

All cosmological hydrodynamical simulations that we present in this work were performed with modified versions of the \textsc{p-gadget3} code, itself an updated and extended version of
\textsc{p-gadget2}\footnote{\url{https://wwwmpa.mpa-garching.mpg.de/gadget/}} \citep{Gadget2}. The code follows the gravitational interactions with
an efficient, parallel, Tree-PM gravity solver and the hydrodynamics with an energy- and entropy-conserving
smoothed-particle hydrodynamics (SPH) scheme \citep{EntropyConservingSPH}. 

The treatment of radiative cooling assumes a primordial composition of the gas with a hydrogen and
helium mass fraction of 76 and 24 per cent respectively. The ionization and thermal state of the gas
is then followed with a non-equilibrium solver that integrates the ionization, recombination,
cooling and heating rate equations using sub-cycling and adaptive time steps \citep[see][]{PuchweinEtAl2015}.
The \textsc{cvode} library\footnote{\url{https://computing.llnl.gov/projects/sundials}} is used for this purpose.
Following the full non-equilibrium equations avoids an artificial delay between 
photoionization and photoheating that is present in simulations with an equilibrium solver \citep[see also][]{Gaikwad2019,Kusmic2022}. 
The following rate coefficients are assumed: the case A recombination rates of \citet{VernerFerland1996},
the \heii\ dielectronic recombination rate of \citet{AldrovandiPequignot1973}, the collisional excitation rates
of \citet{Cen1992}, the collisional ionization rates of \citet{Voronov1997}, and the free-free Bremsstrahlung rate of
\citet{TheunsEtAl1998}. In most of our runs, photoionization and photoheating is followed based on external, spatially homogeneous
models of the UV background (UVB), while in our patchy reionization simulations we interpolate from maps of the radiation field
obtained with the \textsc{aton} radiative transfer code. We will discuss this in more detail below.

Since we are primarily interested
in the low-density IGM, we accelerate our simulations by converting all gas particles that exceed a density
of $1000$ times the mean cosmic baryon density and have a temperature smaller than $10^5\rm\,K$ to collisionless star particles.
While this approach does not yield realistic galaxies, it accurately predicts the properties of the low-density IGM \citep{QuickLya}.

\subsection{Simulations with a homogeneous UVB}
\label{sec:sims_homo_uvb}

For our baseline simulations, we use time-varying but spatially homogeneous photoionization and photoheating rates from
the \textit{fiducial} UVB model presented
in \citet[][see their table D1]{PuchweinEtAl2019}. These were derived in such a way that gas exposed to this UVB follows
a (largely) realistic cosmic reionization and heating history. Hydrogen reionization finishes at $z \sim 6$, while \heii\
reionization ends at $z \sim 2.8$. In particular, simulations with this model avoid an artificially accelerated reionization \citep[see also][]{onorbe2017}.
In comparison, in simulations with, e.g., an \citet{HaardtMadau2012} UVB model, hydrogen reionization would be essentially
completed by $z \sim 11$ \citep[see][]{PuchweinEtAl2019}.

In addition to our baseline simulations, we perform simulations with various modifications of the \citet{PuchweinEtAl2019}
\textit{fiducial} UVB model in order to sample different reionization histories as well as a wider range of IGM temperatures (see \citealt{VillasenorEtAl2022} for a similar approach).
For example, we produce simulations with a colder or hotter IGM by rescaling the photoheating rates while keeping the
photoionization rates fixed. The rescaling factors used for the different runs are provided in Table~\ref{tab:sims}.
For the colder (hotter) models, we use the same factor of 0.5 (2) for the \hi\ and \hei\ photoheating rates,
but a different factor of 0.66 (1.5) for \heii\, as the latter primarily
affects the thermal history at lower redshifts during the epoch of \heii\ reionization.

In addition, we also vary the global reionization redshift (which we define as the time when the volume-averaged neutral fraction in the simulation falls below $10^{-3}$) while keeping the instantaneous gas temperature at $z<5$ fixed
by performing a linear redshift rescaling of our fiducial UV background model \citep{PuchweinEtAl2019} at $z>5$.
The different homogeneous UVB models considered here complete reionization in the redshift range $z_\textrm{r}=5.3$ to 7.4, providing a range of models with different amounts of pressure smoothing. These reionization histories corresponds to a rescaling of the fiducial
UV background redshift coordinate at $z>5$ (i.e. of $z-5$)  by a factor of 
0.89 to 1.24. In addition, to ensure the instantaneous gas temperatures of the models
remain the same at $z<5$, we multiply the \hi\ photoheating rates
by factors 0.9 to 1.5 at $z>5$, with the higher factors being used for models with earlier reionization.

\subsection{Hybrid, patchy reionization simulations}
\label{sec:sims_hybrid_method}

A time-varying but spatially homogeneous UVB model cannot
capture the large spatial fluctuations in the ionizing
radiation field that are present during the patchy cosmic
reionization process. A homogeneous UVB model can only aim to provide
suitable mean values. However, the radiation field will be entirely different
in ionized bubbles and neutral regions. In addition to the
direct effect this has on the ionization state, this also seeds fluctuations
in the IGM temperature and pressure smoothing on large scales.
These fade only slowly and persist well into the post-reionization
epoch \citep[see, e.g.,][]{DAloisioEtAl2015, KeatingEtAl2018}.

To capture these effects, while avoiding the enormous computational
cost of fully coupled radiation hydrodynamics simulations, we
use a new hybrid scheme that combines relatively cheap post-processing
radiative transfer simulations on a fixed Eulerian grid,
with cosmological hydrodynamical simulations that capture
the hydrodynamic response to photoheating by an inhomogeneously evolving
UV radiation field. Our scheme has some common features with the hybrid technique introduced in \citet{OnorbeEtAl2019}. Their scheme uses a semi-numerical, excursion-set method to obtain a map of the local reionization redshift across the simulation volume. For each resolution element in the simulation, they then switch on a time-dependent UVB model at the pre-computed local reionization redshift of the element. The UVB seen by their simulation is otherwise homogeneous, i.e. spatially constant across ionized regions. Our scheme instead uses post-processed radiative transfer simulations to provide an inhomogeneously evolving UV radiation field. This captures the effect of a spatially-varying reionization redshift, but also of spatial fluctuations in the radiation field within ionized regions, which can be significant near the tail end of reionization.

Our scheme consists of the following steps:
\begin{itemize}
\item Performing a cosmological hydrodynamical simulation with a
homogeneous UVB model. We use our baseline simulation discussed above for
this purpose. To provide sufficient time resolution for the next step,
we have saved outputs every 40 Myrs.
 
\item Performing a post-processing radiative transfer simulation on the
outputs of the cosmological hydrodynamical simulation. We do this on
a fixed Eulerian grid with a slightly modified version of the highly efficient, GPU-accelerated
\textsc{aton} code \citep{Aton, Aton2010}. It uses a moment-based radiative
transfer scheme that assumes the M1 closure approximation and uses the full
physical speed of light. We use a number of grid cells that equals
the (initial) number of gas particles in the hydrodynamical simulation
(i.e., $2048^3$ for the patchy simulations presented later in this work).
The challenging task of accurately predicting source luminosities from
galaxy formation physics is bypassed by empirically calibrating the amount of
ionizing radiation escaping from halos to observational constraints on
the ionization state of the IGM, e.g., based on the very high-redshift
\lya\ forest and Thomson scattering optical depth measurements from CMB data. 
Halos are then populated with ionizing sources based on their mass.
Full details of this method are provided in \citet{KulkarniEtAl2019}. For our hybrid simulations completing reionization at a redshift of 5.3, we
use the redshift evolution of the ionizing emissivity that was derived in
that study (see their fig.~1). For our 40 cMpc/$h$ box, we apply a redshift-independent boost factor to the emissivity of $1.265$ to
account for the smaller box size and higher numerical resolution, which changes the resolved
part of the escape fraction. We use a single frequency bin and assume a somewhat smaller (mean) photon energy of $18.63$ eV. We apply the same strategy for calibrating the redshift evolution of the emissivity for our earlier reionization scenarios.

\item The modified \textsc{aton} code version saves maps of the photoionization rate every 40 Myrs. In addition,
it produces a map of the local reionization redshift, which for each grid cell we define as the redshift at which
an ionized, i.e. \hii , fraction of 3 per cent is exceeded for the first time. This records when a cell starts to get
ionized. As cells ionize rapidly once they are reached by an ionization front, the recorded ionization redshift is rather insensitive to the exact threshold value used. Fig.~\ref{fig:zreion_map} shows a slice of a local reionization redshift map produced in this way. Typically high-density regions containing many ionizing sources reionize first, while remote voids are the last regions to reionize.

\begin{figure}
  \centering
  \includegraphics[width=\columnwidth]{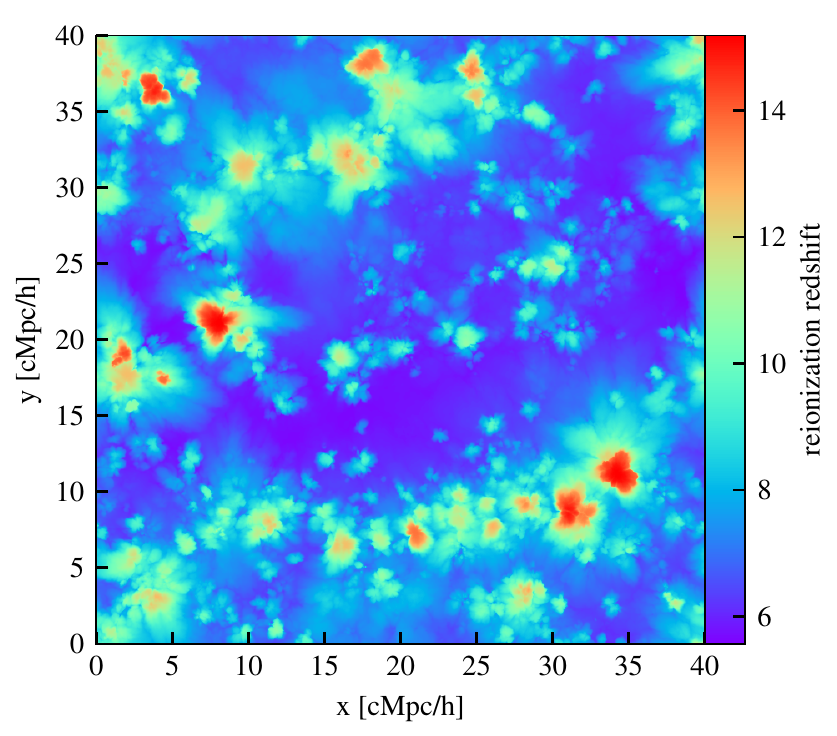}
  \caption{Map of the local reionization redshift in a thin slice produced by the radiative transfer calculation with the \textsc{aton} code. This map corresponds to the $z_\textrm{r}=5.3$ patchy simulation that is discussed later in the manuscript. For comparison, the large scale structures in the same slice are shown in Fig.~\ref{fig:slices_homo_aton_patchy}.}
  \label{fig:zreion_map}
\end{figure}

\begin{figure*}
  \centering
  \includegraphics[width=\linewidth]{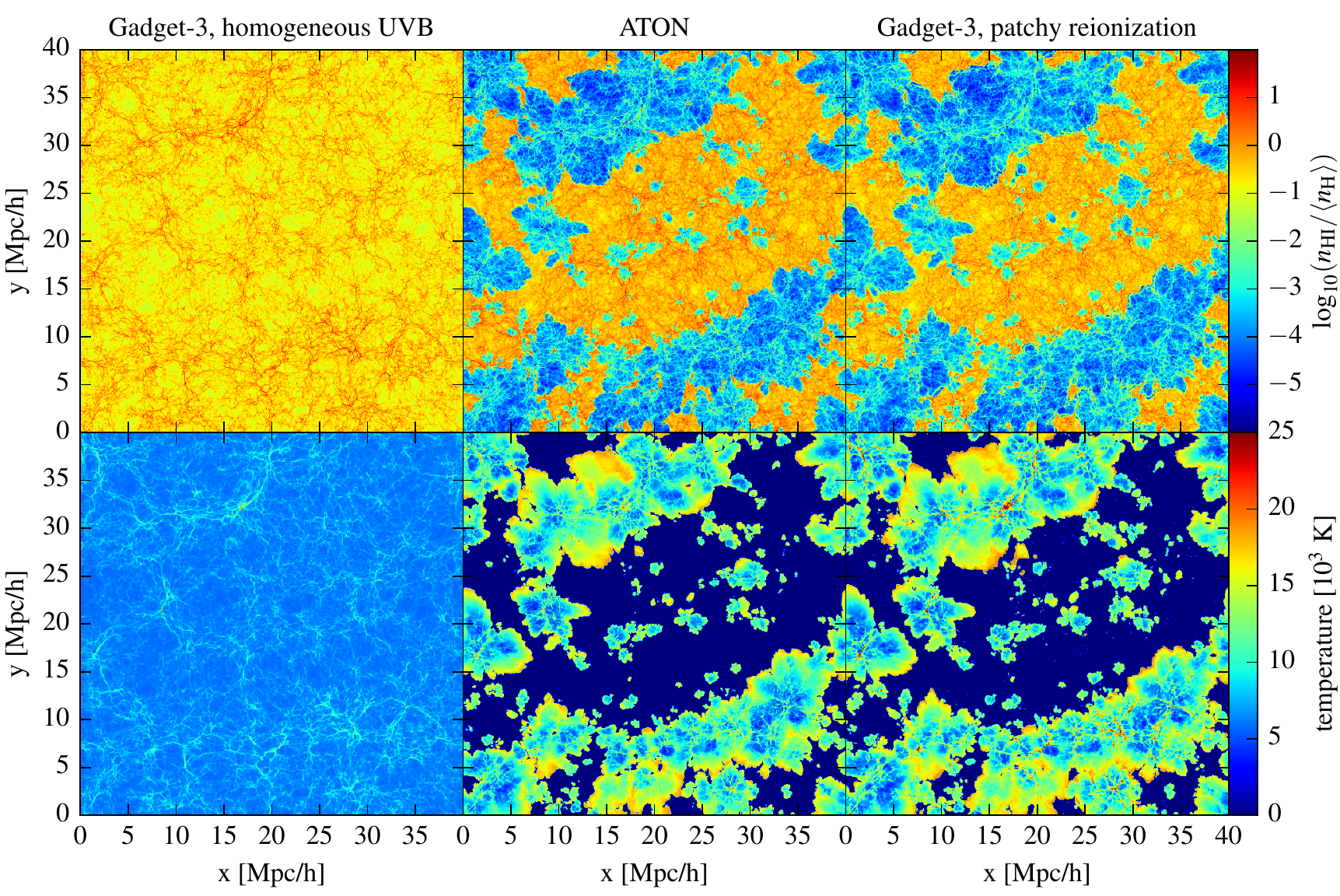}
  \caption{Neutral hydrogen density (in units of the mean total hydrogen density, top panels) and temperature (bottom panels) in a \textsc{p-gadget3} simulation
  with a homogeneous UVB (left, showing the matched homogeneous run for the $z_\textrm{r} = 5.3$ patchy simulation), in an \textsc{aton} post-processing radiative transfer simulation (middle), and in our hybrid, patchy reionization
  cosmological hydrodynamical simulation (right). Results are shown at $z\sim7$ for simulations that complete reionization at $z_\textrm{r} = 5.3$. 
  By construction, all three simulations have very similar (volume-weighted) \hi\ fractions of $\sim 45\%$ and temperatures at mean density of $T_0 \sim 6900$ K.
  While the hybrid simulation has similar ionized/neutral regions as the \textsc{aton} run, it also accounts for 
  shock heating of gas in high-density regions, consistent pressure smoothing, and the advection of the thermal energy in gas flows. 
  }
  \label{fig:slices_homo_aton_patchy}
\end{figure*}

\item Finally, we perform a second cosmological hydrodynamical simulation of the same volume. In this run
our modified version of \textsc{p-gadget3} loads the maps of the photoionization rate produced by \textsc{aton}
and uses them as a spatially varying UV background for following photoionization and photoheating
during patchy reionization. At each time step and for each SPH particle, local values of the the photoionization and
photoheating rates are computed as follows. If the host cell of the particle has not started to be photoionized in the \textsc{aton}
simulation yet (\hii\ fraction continues to be smaller than 3 per cent), the rates are assumed to be zero so that we can
completely skip the integration of the rate equations and assume the gas in this SPH particle is neutral. At redshifts lower
than the local reionization redshift, the \hi\ photoionization rate of the host cell is interpolated between the nearest \textsc{aton} output
times and adopted for the SPH particle and the current time step. A slightly different treatment is used right after an ionization front has reached a particle, i.e.
between the local reionization redshift assigned from the map and the next (lower) \textsc{aton} output redshift. In this case we use the rate of the next (lower redshift) map directly without interpolation. This results in a larger jump in the photoionization rate at the recorded local reionization redshift, corresponding to a quickly passing ionization front.
Our tests showed that this leads to a smoother growth of the ionized regions in the hydrodynamical simulation, thereby further reducing residual imprints
of the finite number of \textsc{aton} outputs. Since we cannot capture differences between \hi\ and \hei\ reionization with
a single frequency bin in the radiative transfer, we simply adopt the \hi\ photoionization rate also for \hei.
The \hi\ photoheating rate is computed based on the assumed mean photon energy, i.e. the photoionization rate
is simply multiplied by $18.63 \, \textrm{eV} - 13.6 \, \textrm{eV} = 5.03 \, \textrm{eV}$. The \hei\ photoheating
rate (per atom) is then assumed to be 1.3 times that of \hi, in rough agreement with the ratio of the two in
the homogeneous, synthesis UVB model from \citet{PuchweinEtAl2019}. Finally, the \heii\ photoionization and photoheating rates are assumed to be
spatially homogeneous and are adopted from the \citet{PuchweinEtAl2019} \textit{fiducial} UVB model. These latter rates play, however,
a role only at lower redshifts and are negligible during the epoch of \hi\ reionization, except in the proximity of quasars \citep[e.g.][]{Bolton2012}, which we do not model here. 
Our hybrid technique could be extended to lower redshifts by using multi-frequency radiative transfer simulations that follow \heii\ reionization, alternatively a semi-analytic model like in \citet{2020MNRAS.496.4372U} could be used. Here, we refrain from such attempts, focus on high redshifts, and use the local rates derived as described above to integrate
the ionization and cooling/heating rate equations in the same manner as in our homogeneous UVB simulations.
\end{itemize}

The results from a patchy, hybrid radiative transfer/cosmological hydrodynamical simulation are shown at
redshift $z\sim7$ in Fig.~\ref{fig:slices_homo_aton_patchy}. The left-hand panels show the neutral hydrogen density
(top) and gas temperature (bottom) in a \textsc{p-gadget3} simulation with a homogeneous UVB.
The middle panels show the post-processed radiative transfer simulation performed with the \textsc{aton} code,
while the right-hand panels show the results of the hybrid radiative transfer/cosmological hydrodynamical simulation.
As expected, the simulation with a homogeneous UVB completely misses the patchy nature of cosmic reionization.
The \textsc{aton} simulation nicely displays the complicated morphology of ionized bubbles, but misses some
important physics. In particular, running the calculation in post-processing means that the dynamical effect of
inhomogeneous photoheating (e.g., pressure-smoothing) and its impact on the gas density distribution cannot be followed.
In addition, the thermal energy injected by photoheating is stored for each grid cell, but not properly advected with
the gas flow. This can sometimes be seen as dense gas leaving a wake of increased temperature when it falls towards
a structure. Also the temperature evolution in the \textsc{aton} simulation only accounts for photoheating and
adiabatic evolution, but misses shock heating of gas in dense regions. The patchy, hybrid \textsc{p-gadget3} 
cosmological hydrodynamical simulation (right panels) captures all these aspects. The morphology of ionized and
neutral regions is almost identical to the \textsc{aton} simulation, but, e.g., the temperature in halos is larger
due to the inclusion of shock heating. The dynamical impact of inhomogeneous photoheating in our patchy \textsc{p-gadget3} 
simulation will be discussed in detail in Sec.~\ref{sec:pressure_smooth}.

At first glance our multi-step hybrid radiative transfer/cosmological hydrodynamical simulation scheme may seem complicated, e.g.,
compared to a single fully coupled radiation-hydrodynamics simulation. There are, however, several distinct
advantages. First, the computational cost is much lower compared to a full radiation hydrodynamics calculation (e.g., 0.5 million core hours on CPUs + 3 thousand GPU hours for our $z_\textrm{r}=5.3$ patchy run compared to 28 million core hours on CPUs for the main Thesan run; \citealt{Thesan}).
The reason for this is that the radiative transfer is done on a somewhat coarse fixed Eulerian grid (cell size $19.5 \, h^{-1} \, \textrm{ckpc}$ in our $40 \, h^{-1} \, \textrm{cMpc}$
patchy simulations), and hence allows for relatively large Courant time steps as well as an efficient parallelization on 
powerful GPUs. Second, as we empirically calibrate the emission that escapes into the IGM, we can continue to use our strongly simplified
galaxy/star formation model (see Sec.~\ref{sec:sim_code}) and avoid all the complications of radiation-hydrodynamically modelling realistic
source galaxy populations as well as the escape of ionizing radiation from them. Third, finding a calibration of the source
luminosity as a function of halo mass that agrees with observational constraints on the ionization state of the IGM (using only cheap post-processing
radiative transfer simulations for this purpose) is much simpler than modifying a full simulation model of galaxy formation in such a way that the same
is achieved. Finally, the overhead of producing the first hydrodynamical simulation with a homogeneous UVB came (at least in this project) for free in practice,
as we would have needed this run in any case as a baseline model for the comparison to the large number of simulations with different ionization histories and dark
matter models (see Sec.~\ref{sec:relics_suite}).

Of course the computational efficiency of our approach also comes at a price. By saving only a limited number of maps of the radiation field on a fixed grid,
we have access to the radiation field only with a limited spatial and time resolution when following photoionization and photoheating in
the second hydrodynamical simulation. Also, the coupling of the radiative transfer to the hydrodynamics is not fully self-consistent in cases where the difference in
pressure-smoothing between the first and second hydrodynamic simulation significantly changes the opacity of the medium and the local radiation field.
We expect these effects to primarily play a role on small scales in dense systems. We would hence not advise to use this method for investigating, e.g., the escape of ionizing
radiation from galaxies, and one should probably be careful when studying the details of self-shielding of dense gas. The impact of large scale fluctuations
(on the scale of the size of ionized bubbles) on ionization and pressure-smoothing of the low-density IGM should, however, be robustly predicted.  

\begin{figure}
  \centering
  \includegraphics[width=\columnwidth]{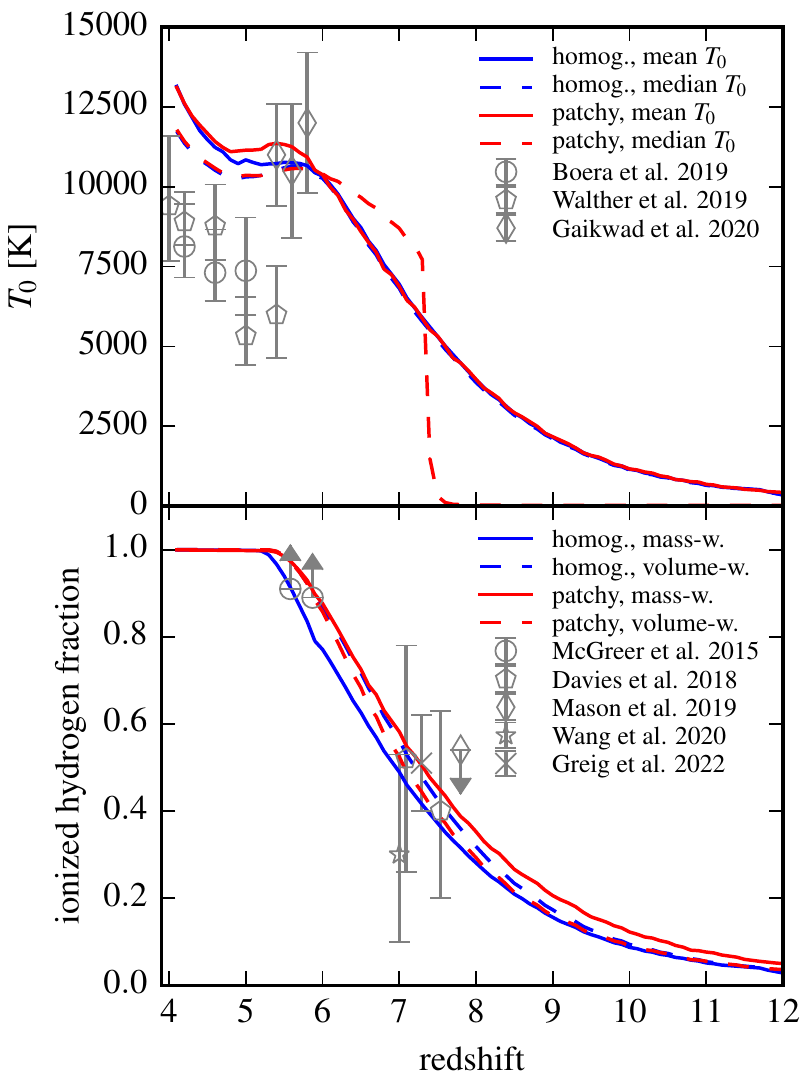}
  \caption{IGM properties as a function of redshift in the patchy simulation that completes reionization at $z_\textrm{r}=5.3$, as well as in the matched homogeneous simulation. The temperature at mean baryon density ($T_0$) is shown in the top panel. Results are indicated for the volume-weighted mean and median value of $T_0$, measured in gas with densities $0.975 < \Delta < 1.025$. IGM temperature measurements from \citet{BoeraEtAl2019}, \citet{2019ApJ...872...13W} and \citet{2020MNRAS.494.5091G} are shown for comparison. The bottom panel shows the mass- and volume-weighted ionized fractions (i.e. the \hii\ fractions). Observational constraints from \citet{2015MNRAS.447..499M}, \citet{2018ApJ...864..142D}, \citet{2019MNRAS.485.3947M}, \citet{2020ApJ...896...23W} and \citet{2022MNRAS.512.5390G} are shown for reference.}
  \label{fig:temp_xhii_taueff_vs_z}
\end{figure}

\begin{figure*}
  \centering
  \includegraphics[width=\linewidth]{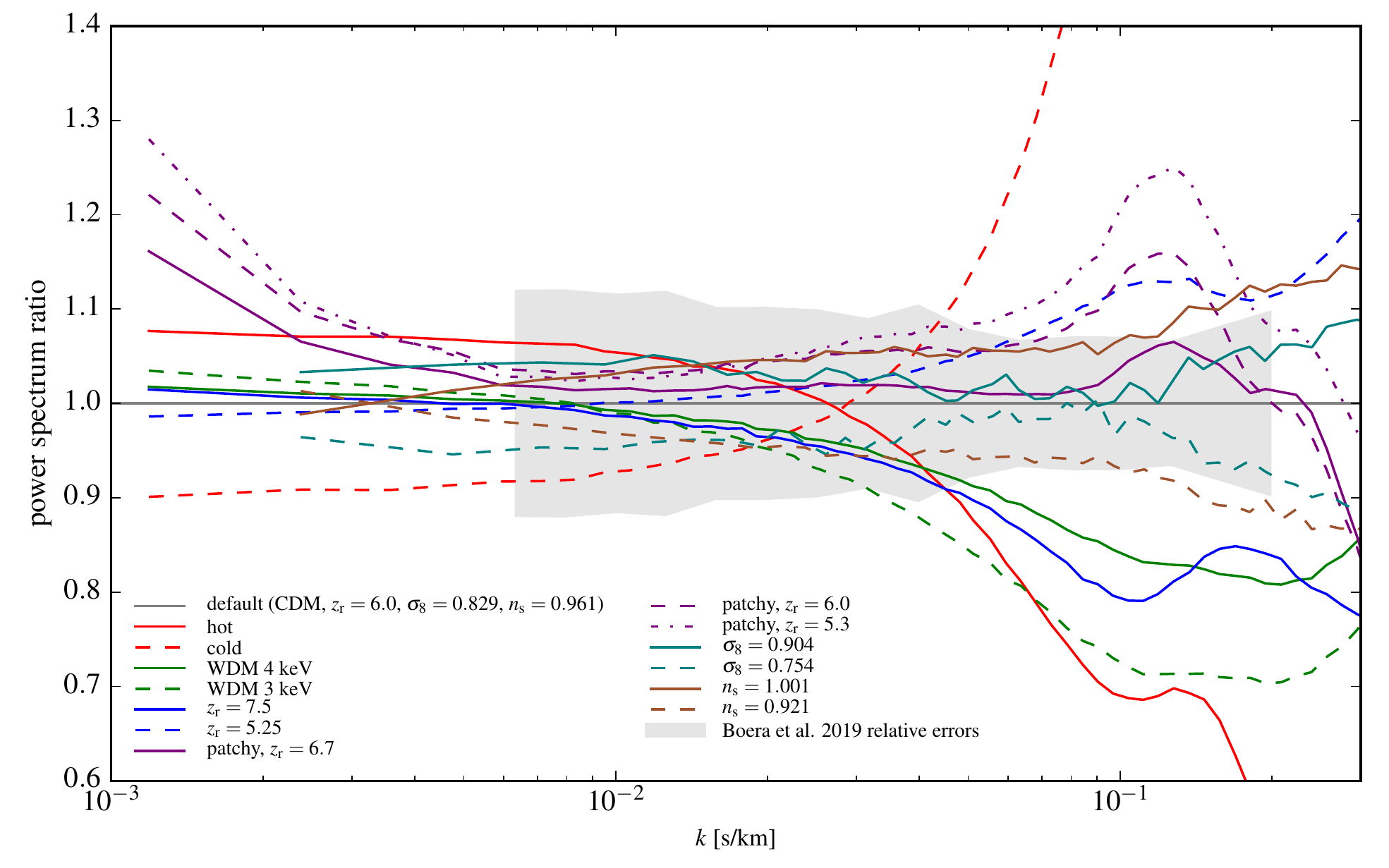}
  \caption{Change in the one-dimensional \lya\ forest flux power spectrum due to different choices of reionization/photoheating model, dark matter model, as well as matter power spectrum normalization and slope. Shown are the ratios of the \lya\ forest flux power spectrum in the considered simulations to that of our default (or baseline) $\Lambda$CDM model with a \citet{PuchweinEtAl2019} UVB at $z=4.6$. In the hot/cold models, the \hi\ photoheating rate has been increased/decreased by a factor of 2. For reference, the shaded region indicates the relative error in the power spectrum measurement of \citet{BoeraEtAl2019}. Significant degeneracies between different modifications exist. Interestingly, the increase in power on large spatial scales in our patchy simulations seems to be a characteristic signature of inhomogeneous reionization.
  }
  \label{fig:pow_spec_ratios_grid}
\end{figure*}

To isolate the effects of the patchiness of reionization from the effects of differences in the reionization history,
we have performed simulations with homogeneous UVB models that produce the same 
average reionization and thermal histories as our hybrid, patchy reionization simulations. Details on how a suitably tailored UVB model for such a simulation
is obtained are provided in Appendix~\ref{app:homogeneous_UVB}. We have performed such pairs of patchy and matched homogeneous simulations for different reionization histories with reionization completing at redshifts $z_\textrm{r} = 5.3$, 5.7, 6.0 and 6.6. We will mostly concentrate on the first model in the analysis as its reionization history seems in best agreement with observations \citep[e.g.,][]{KulkarniEtAl2019,bosman2022}. 

Fig.~\ref{fig:temp_xhii_taueff_vs_z} compares this matched homogeneous model to the corresponding patchy simulation. It displays the evolution of the mean and median IGM temperature at mean density, as well as the mass- and volume-weighted ionized hydrogen fraction. As planned, the IGM temperature and ionized fraction in the matched homogeneous run closely follow those in the patchy simulation. We have opted to follow the mean temperature at mean density of the patchy simulation during reionization, and the median temperature at mean density after reionization (see Appendix~\ref{app:homogeneous_UVB} for further details on this). 

Note that, despite the similar neutral fractions, the \lya\ forest transmission properties will be quite different during reionization. While ionized regions can allow transmission in a patchy reionization scenario, a small amount of homogeneously distributed residual neutral gas that is present even late in the reionization process is sufficient to almost fully absorb the \lya\ forest in a homogeneous model. This too homogeneous distribution of neutral gas, which can be seen in the upper, left panel of Fig.~\ref{fig:slices_homo_aton_patchy}, is the main reason why simulations with a homogeneous UVB cannot reproduce the observed statistics of the \lya\ forest at very high redshift ($z \gtrsim 5.3$; e.g., \citealt{BeckerEtAl2015}). Patchy reionization simulations, even when done in post-processing, do a much better job \citep[e.g.,][]{KulkarniEtAl2019, KeatingEtAl2020, bosman2022}.

\section{Results}

\subsection{The impact of cosmology, reionization model and IGM temperature on the \lya\ forest}
\label{sec:grid_pow_spec}

The \lya\ forest is sensitive to a wide range of cosmological and astrophysical parameters and processes. We aim to sample many of the most relevant ones with the Sherwood-Relics simulation suite. Fig.~\ref{fig:pow_spec_ratios_grid} provides an overview of the various impacts on the structures present in the \lya\ forest on different scales. The figure shows the relative change in the one-dimensional \lya\ forest flux power spectrum compared to our baseline simulation with cold dark matter and a \citet{PuchweinEtAl2019} UVB.

Note that by $z=4.6$, even in our patchy reionization simulations, large scale fluctuations in the ionizing radiation field have largely faded. We thus choose to rescale the optical depths in all simulations shown in the figure such that the mean transmission value is consistent with observations at that redshift. We have used the following fitting function for the observed effective optical depth for this purpose \citep[see][]{MolaroEtAl2022},
\begin{equation}
\tau_\textrm{eff} =
\begin{cases}
   -0.132 + 0.751 \, [(1 + z)/4.5]^{2.90},& \text{if } 2.2 \leq z < 4.4\\
   1.142 \, [(1 + z)/5.4]^{4.91}. & \text{if } 4.4 \leq z \leq 5.5.
\end{cases}
\label{eq:tau_eff_fit}
\end{equation}

Fig.~\ref{fig:pow_spec_ratios_grid} displays ratios of the power spectrum of the normalized transmitted flux contrast, i.e. of $F/\bar{F}-1$, where $F$ is the normalized transmitted flux and $\bar{F}$ denotes its mean value. For comparison, the relative errors in the \citet{BoeraEtAl2019} \lya\ forest flux power spectrum measurement are also indicated.

\begin{figure*}
  \centering
  \includegraphics[width=\linewidth]{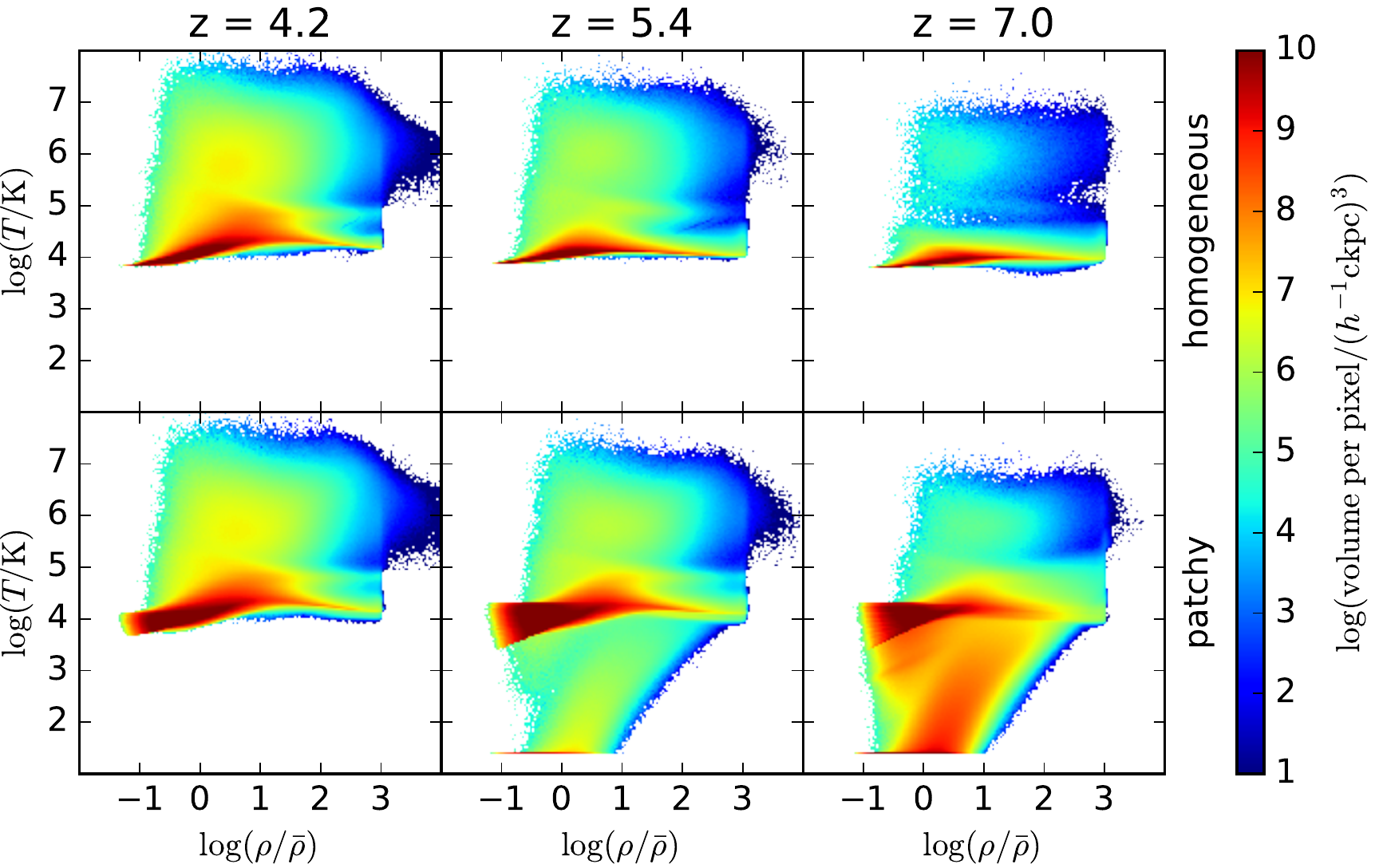}
  \caption{Volume-weighted temperature-density distribution of gas in the patchy simulation that completes reionization at $z_\textrm{r}=5.3$ (bottom panels) and in the matched homogeneous simulation (top panels) at redshifts $4.2$, $5.4$ and $7.0$. Both simulations have, by construction, similar volume-weighted neutral fractions of $\sim$0\%, 4\%, 45\% at these redshifts, respectively. During reionization, cold gas that has not yet been ionized is present in the patchy run. In contrast, the same gas is partly ionized and heated in the homogeneous simulation. After reionization, the temperature-density relation in the patchy simulations is still broader at low densities.
  }
  \label{fig:rho-T}
\end{figure*}

Clearly, the different changes in the simulated physics leave specific imprints in the flux power spectrum. Models with a hotter/colder IGM (with \hi\ and \hei\ photoheating rates boosted/reduced by a factor of 2, see Sec.~\ref{sec:sims_homo_uvb} for full details) have significantly less/more power on small spatial scales (large $k$). This is consistent with the expectation of increased thermal broadening and pressure smoothing at higher temperature. Pressure smoothing is also increased by an earlier reionization (see the $z_\textrm{r}=7.5$ model, blue solid curve). Interestingly, the corresponding suppression of power on intermediate scales, $10^{-2}\,\textrm{s}\,\textrm{km}^{-1} \lesssim k \lesssim 10^{-1}\,\textrm{s}\,\textrm{km}^{-1}$, is very similar to that in a warm dark matter model with 4 keV particle mass (green solid curve). This already suggests that there will be degeneracies between dark matter constraints and the thermal/reionization history of the IGM  \citep[see e.g.][]{Viel_2013,2017PhRvD..96b3522I,Garzilli2019}. One can aim to break these by including measurements at different redshifts, as well as at higher $k$ \citep[e.g.][]{Nasir2016}. Changing the cosmological parameters $\sigma_8$ and $n_\textrm{s}$ has largely the expected effects \citep[see also][]{QuickLya,McDonald2005}. Similar to the matter power spectrum, it changes the normalization and slope of the \lya\ forest flux power spectrum. It is worth noting that the simulations with different dark matter particle masses and cosmological parameters shown in Fig.~\ref{fig:pow_spec_ratios_grid} use the same UVB/reionization model. This, hence, isolates the direct impact of these parameters on the IGM from their effects on the ionizing source galaxy population. There would be additional effects, in particular during reionization, when also modelling the impact of cosmology on the ionizing sources and hence reionization history and topology \citep[e.g.,][]{2014MNRAS.438.2664S,2017PhRvD..96j3539L,2021MNRAS.508.1262M}. We (partly) capture these effects by covering a range of reionization redshifts with our sample, so that it can be varied as a separate parameter when doing parameter inference. In this way, we can be agnostic about the details of the impact on the source galaxies.

Furthermore, we find that, as already noted elsewhere \citep{Cen2009,KeatingEtAl2018,DAloisio2018,OnorbeEtAl2019,WuEtAl2019,MonteroCamacho2020,MolaroEtAl2022}, the patchy reionization simulations predict distinctive increases in power on the largest spatial scales (smallest $k$). We will discuss these in detail in Sec.~\ref{sec:modulation_lya}.

Our grid of simulations, as detailed in Table~\ref{tab:sims} and as (partly) shown in Fig.~\ref{fig:pow_spec_ratios_grid}, will be used for parameter inference studies in forthcoming work. In the remainder of this study, we will concentrate on our new hybrid simulations and discuss the \textit{relic} signatures that patchy cosmic reionization leaves in the IGM and \lya\ forest.  

\begin{figure*}
  \centering
  \includegraphics[width=\linewidth]{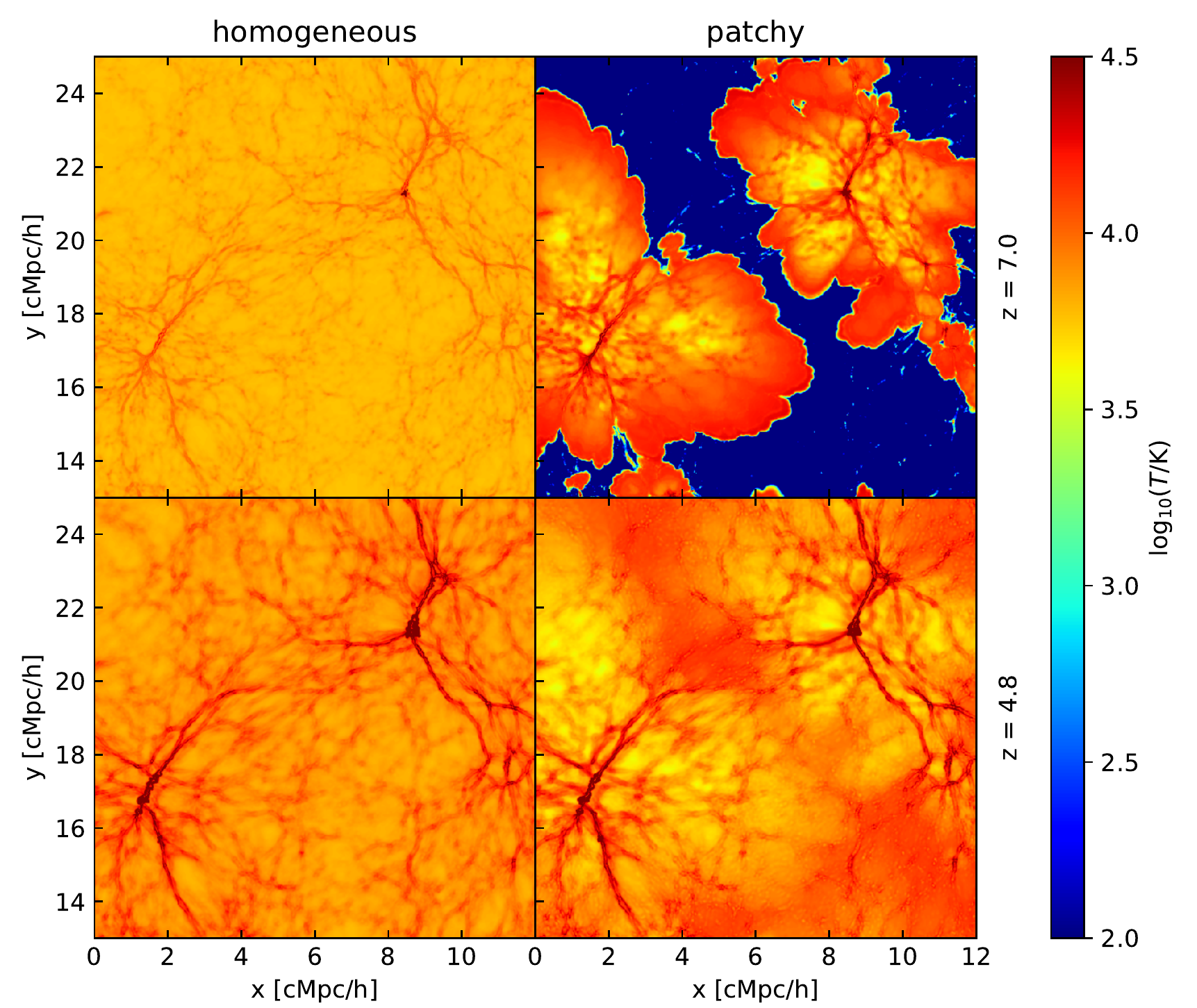}
  \caption{Gas temperature in a thin slice through our matched homogeneous and patchy simulations which complete reionization at $z_\textrm{r}=5.3$. During reionization (redshift 7, both runs $\sim45\%$ neutral fraction, upper panels) the gas temperature in the simulation with the homogeneous UVB has only small fluctuations which largely trace the gas density, as expected from the narrow density-temperature relation shown in Fig.~\ref{fig:rho-T}. In contrast the temperature in the patchy simulation differs by orders of magnitude between ionized and neutral regions. Some of the highest temperatures (outside shock heated regions) are found near the ionization fronts in recently ionized gas which had little time to subsequently cool. After reionization (redshift 4.8, lower panels), the temperature in the homogeneous simulation still looks qualitatively similar (except for more pronounced shock heating than at $z=7$). In the patchy simulation, all of the gas has been photoheated as well, but large scale temperature fluctuations that are relics of the patchy reionization process are still clearly visible. The highest temperatures (except for shock heated gas) are found in regions that have been reionized late.}
  \label{fig:slice_temp}
\end{figure*}

\subsection{The impact of patchy reionization on the IGM and the \lya\ forest}
\label{sec:patchy_results}
\subsubsection{The thermal state of the IGM during and after patchy cosmic reionization}
\label{sec:thermal_properties}

During the era of cosmic reionization, energetic UV photons emitted by first galaxy populations ionize the IGM. The excess energy of these photons beyond the ionization energy of the relevant atoms/ions is available for heating the IGM. Given the patchy nature of cosmic reionization, this causes significant temperature differences between regions that reionize at different times.

Fig.~\ref{fig:rho-T} shows how this affects the temperature-density distribution of the IGM during and after cosmic reionization. At $z=7$, when roughly half of the hydrogen is ionized, there is both cold, neutral and hot, ionized gas present in the $z_\textrm{r}=5.3$ patchy simulation, corresponding to neutral and ionized regions. The matched homogeneous UVB simulation instead contains only gas that has a temperature of at least several thousand Kelvin. Except for a small amount of shock-heated gas, the gas is partly ionized and partly photoheated (see also Fig.~\ref{fig:slices_homo_aton_patchy}) and follows a tight temperature-density relation that is almost flat. This corresponds to all gas having a similar thermal history with little variation around the mean evolution shown in Fig.~\ref{fig:temp_xhii_taueff_vs_z}. It also explains the very similar mean and median thermal evolutions in the homogeneous model that are also indicated there. In the patchy simulations instead, there is more variation in temperature of the ionized gas, in particular at low densities where temperatures span a range of 3000 - 20000 K, corresponding to different local reionization redshifts and consequently different amounts of cooling after reionization \citep[e.g.,][]{2007MNRAS.380.1369T, 2008ApJ...689L..81T}. Additional broadening of the temperature-density relation is expected from the evaporation of small structures by photo-heating which causes their gas content to cool by adiabatic expansion while driving shocks into small nearby voids that are thereby heated \citep{2018MNRAS.474.2173H}. The situation at $z=5.4$ is qualitatively similar, just with much less cold, neutral gas remaining in the patchy run.

\begin{figure*}
  \centering
  \includegraphics[width=\linewidth, trim=0 0 1cm 0, clip]{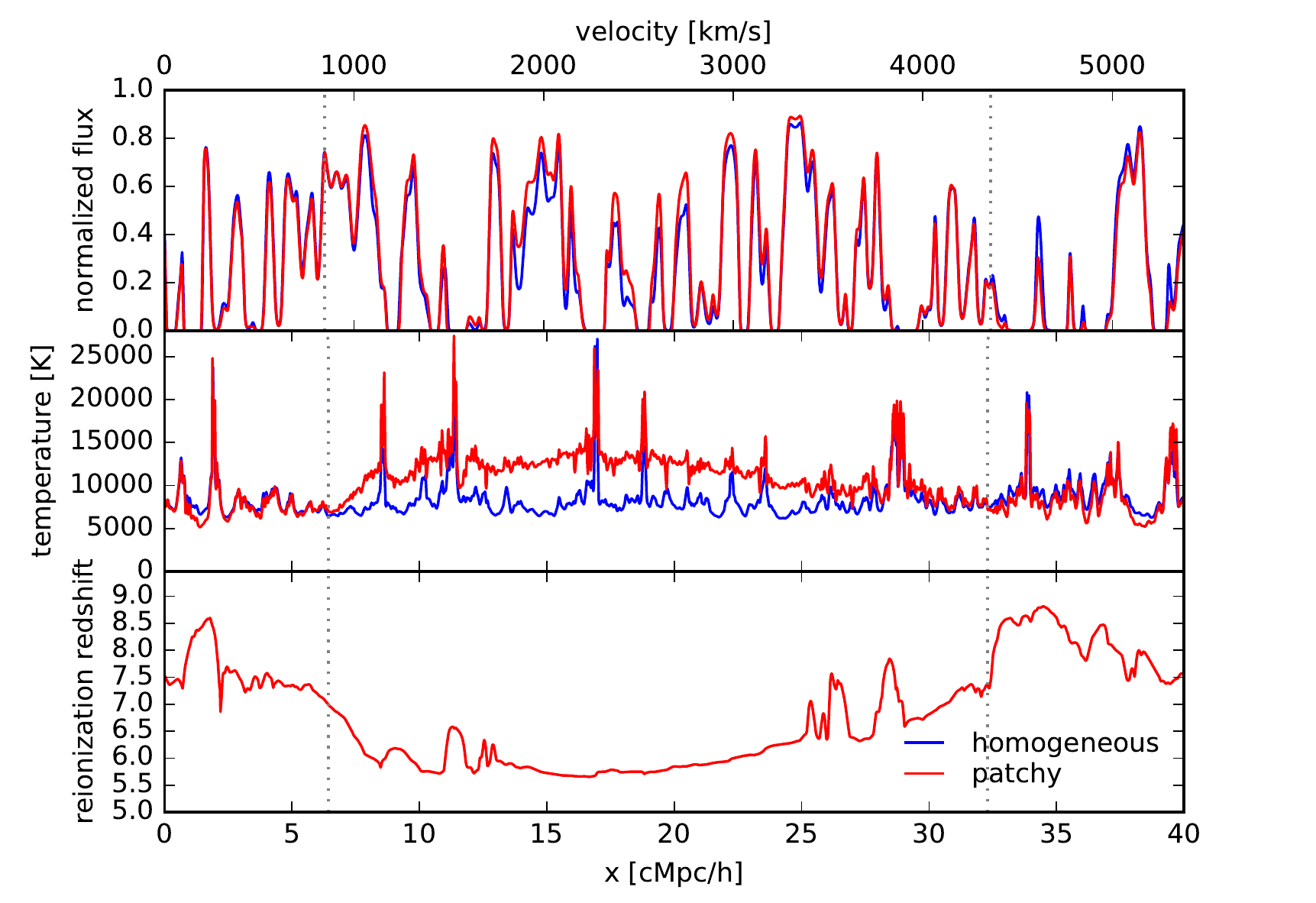}
  \caption{Impact of patchy reionization on the \lya\ forest at $z=4.8$. The top panel displays the normalized transmitted \lya\ fluxes along the same line-of-sight through the patchy (red) and matched homogeneous simulations (blue). The optical depths were re-scaled to be consistent with the observed mean transmission (see Eq.~\ref{eq:tau_eff_fit}). In the left and right regions (roughly left of the first and right of the second vertical dotted line), there is slightly less transmission in the patchy run, while it exhibits more transmission than the matched homogeneous run in the central region. The middle panel shows that this finding is closely related to large scale temperature fluctuations in the patchy simulation in which the central region has a higher temperature and is hence more highly ionized. The temperature fluctuations are in turn largely driven by spatial variations of the local reionization redshift (bottom panel). Regions that have been reionized late are typically hotter.}   
  \label{fig:specs_homog_patchy}
\end{figure*}

At $z=4.2$, well after the end of reionization (at $z_\textrm{r}=5.3$), the homogeneous and patchy simulations look more similar. The temperature-density relations have steepened somewhat (at least the upper envelope for the patchy run). In ionization equilibrium photoionizations balance recombinations. Recombinations happen more frequently in dense gas which consequently receives more photoheating per particle. In addition, the lowest density gas, i.e. the gas in voids, expands more strongly during cosmic expansion and structure formation resulting in increased adiabatic cooling. In the patchy run, there is nevertheless a larger spread in temperature in very low density gas as the temperature fluctuations seeded by inhomogeneous reionization fade only slowly.

This is also shown in Fig.~\ref{fig:slice_temp}, which displays the IGM temperature in a thin slice through part of the simulation box. During reionization, at $z=7$, the temperature map of the patchy simulation clearly shows the locations of ionized regions which have been strongly heated, while neutral regions are still cold. The effect discussed above that recently heated low density gas is hotter than gas that has been reionized earlier is also visible. It results in the temperatures of gas in ionized regions that is located near the ionization fronts being particularly high. We have already seen from the temperature-density distributions that even at $z=7$ all gas has been heated significantly in the homogeneous UVB run. This is reflected by the temperature map in the upper, left panel of Fig.~\ref{fig:slice_temp} which shows no unheated gas.

At $z=4.8$, i.e. $\Delta z \approx 0.5$ after the end of reionization in these simulations, large scale temperature fluctuations are still present in the patchy run. This relic signature of patchy reionization reflects when regions were reionized. One can see this clearly by comparing the upper and lower right panels of Fig.~\ref{fig:slice_temp}. Regions that were still cold and neutral at $z=7$ are particularly hot at $z=4.8$ as they have been reionized late and had less time to cool after their reionization.

\subsubsection{Modulation of the \lya\ forest on large scales}
\label{sec:modulation_lya}

\begin{figure}
  \centering
  \includegraphics[width=\linewidth]{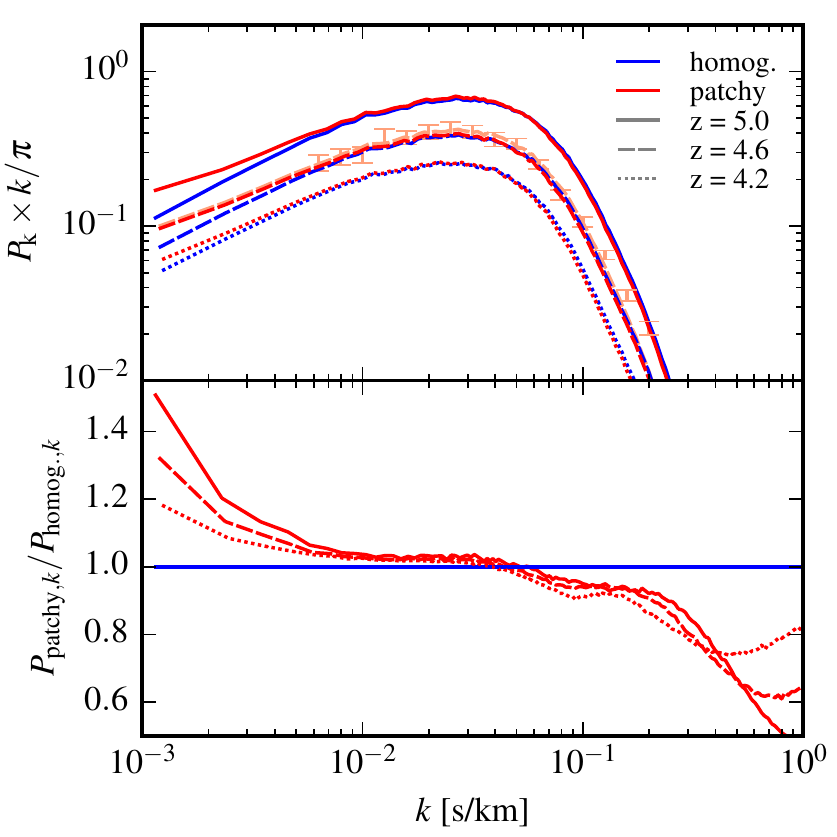}
  \caption{One-dimensional power spectrum of the normalized transmitted \lya\ flux contrast in the $z_\textrm{r}=5.3$ patchy and matched homogeneous simulations (top). The optical depths were re-scaled to be consistent with the observed mean transmission (see Eq.~\ref{eq:tau_eff_fit}). Results are shown at $z=5.0$, $4.6$ and $4.2$. An enhanced large scale (small $k$) power is found in the patchy simulation in particular near the end of reionization. This corresponds to the large scale modulation of the transmission that is displayed in Fig.~\ref{fig:specs_homog_patchy}. 
  For reference, the error bars show resolution-corrected observational constraints from \citet{BoeraEtAl2019} at $z\sim4.6$. They can be compared to the light red, dashed curve, which again shows the patchy simulation at $z=4.6$, but with the mean transmission scaled to the value measured in the observational data, and with a correction for numerical resolution applied. The curve is overall in good agreement with the data, except for the two highest $k$ data points which are most susceptible to differences in the thermal history, as well as to residual metal contamination and resolution effects.
  The bottom panel shows the ratio of the power spectrum in the patchy to that of the homogeneous simulation. In addition to the large scale enhancement, a reduction of power on small scales is visible in the patchy simulation \citep[see also][]{MolaroEtAl2022}.}
  \label{fig:flux_pow_specs}
\end{figure}

These large scale temperature fluctuations also affect the ionization state of the gas. In equilibrium the neutral fraction is proportional to the recombination rate, which depends on temperature and is roughly proportional to $T^{-0.7}$. Hence, hot regions have a lower neutral fraction and correspondingly allow more \lya\ forest transmission. This results in a large scale modulation of the \lya\ transmitted flux.

Fig.~\ref{fig:specs_homog_patchy} illustrates this effect. The bottom panel shows the local reionization redshift along a line-of-sight through the simulation box of the $z_\textrm{r}=5.3$ patchy run. The middle panel displays the IGM temperature along the same skewer at $z=4.8$. Temperatures are shown for both the patchy run and the corresponding matched homogeneous run. The late reionizing region in the middle ($x\approx 6$ to 32 cMpc$/h$) has an increased temperature in the patchy run, as there is little time to cool between its reionization and $z=4.8$. The increased temperature in turn results in a lower neutral fraction and hence more transmission in the \lya\ forest in that region in the patchy run compared to the matched homogeneous run. This is shown in the upper panel of Fig.~\ref{fig:specs_homog_patchy}. In early reionizing regions, the opposite effect can be seen, the transmission is slightly lower in the patchy reionization simulation.

Note that the optical depths in both simulations have been rescaled such that (when averaged over our full sample of 5000 lines-of-sight through the box) they are consistent with the observed mean transmission value (according to Eq.~\ref{eq:tau_eff_fit}). The large scale fluctuation in the transmitted flux is, hence, driven by the temperature variation and not by fluctuations in the radiation field in our patchy simulations, which have largely faded by $z=4.8$.

To summarize, Fig.~\ref{fig:specs_homog_patchy} illustrates how a spatially varying reionization redshift translates to a large scale modulation of the \lya\ forest transmitted flux. Such a modulation is also expected to change the \lya\ forest transmitted flux power spectrum. Fig.~\ref{fig:flux_pow_specs} shows that this is indeed the case. 
Results are indicated for several redshifts after the end of reionization.
Clearly visible is an increased power in the patchy simulation on large scales, $k \approx 10^{-3}$ to a few times $10^{-3}\rm\,s\,km^{-1}$, corresponding to modes with peaks having sizes of $\lambda/2 \gtrsim 5 \, \textrm{cMpc}/h$, which are typical sizes of ionized regions (compare to Fig.~\ref{fig:slices_homo_aton_patchy}). The power is increased all the way up to the fundamental mode of the $40 \, \textrm{cMpc}/h$ box. As expected for an effect caused by large scale temperature fluctuations seeded by patchy reionization, the power enhancement fades away at lower redshift. On small spatial scales, $k \gtrsim 0.05 \rm \, s\,km^{-1}$, there is less power in the patchy simulation compared to the matched homogeneous simulation. This behaviour on large and small spatial scales is consistent with that obtained by \citet{WuEtAl2019} using fully coupled radiation-hydrodynamics simulations that follow the transfer of radiation also with the M1 method. In contrast, and maybe somewhat surprisingly, \citet{2022ApJ...928..174M} do not find a significant upturn of the flux power spectrum on large spatial scales in their simulations of patchy reionization that use the Optically Thin Variable Eddington Tensor technique for the radiative transfer. The origin of this discrepancy is currently unclear.

We find that the behaviour of the power spectrum on small spatial scales (large $k$) is sensitive to the pressure smoothing of the gas and hence requires a coupling of the radiative transfer to the hydrodynamics to be faithfully followed. The increase of power on large spatial scales (small $k$) is in contrast also captured by post-processing radiative transfer simulations \citep[e.g.,][]{KeatingEtAl2018}. The details of this increase may depend on the model of the ionizing source population, which is in turn affected both by the assumed astrophysics and cosmology. We will explore this further in future studies, but \citet{2022ApJ...931...62H} suggest that this may have rather mild impacts on large spatial scales.

Employing several of our patchy reionization simulations, the causes of the reduction of the power spectrum on small scales were investigated in detail by \citet{MolaroEtAl2022}. The main findings were that the spatial fluctuations in the thermal broadening kernel and the different peculiar velocity fields in the patchy simulations cause the reduction in small scale power. For example, transmission spikes will appear first in low density regions that typically reionize late and are hence particularly hot in the patchy model. Consequently, there will be more thermal broadening in these regions, reducing the flux power spectrum on small scales.

Note, however, that the reduction of power on small scales discussed above is based on a matched comparison of a patchy and homogeneous run with the same mean ionization and thermal history. When comparing to a homogeneous run with a different and more extended thermal history, like our baseline run with a \citet{PuchweinEtAl2019} UVB, there can be more power on small scales in the patchy simulation (compare to Fig.~\ref{fig:pow_spec_ratios_grid}).

\begin{figure*}
  \centering
  \includegraphics[width=\linewidth]{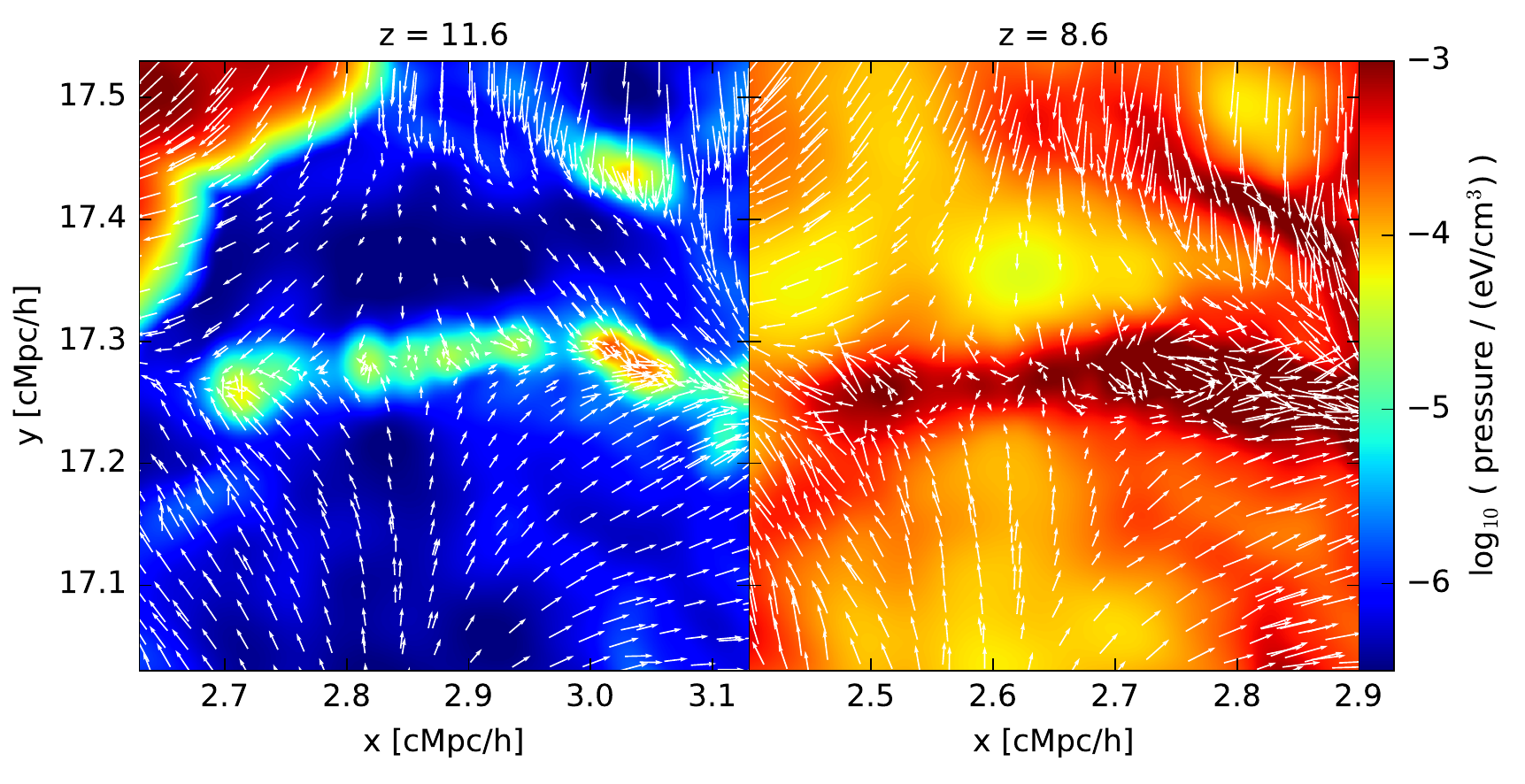}
  \caption{Gas pressure in thin slices through the $z_\textrm{r}=5.3$ patchy simulation. Results are shown at $z\approx11.6$ (left panel), when the region is still largely neutral (only the upper left corner has already been swept over by the approaching ionization front), and at $z\approx8.6$ (right panel), after the region has been reionized. Both panels show the same structures, although note that different $x$-coordinate ranges have been used because the whole region is falling (almost exactly) towards the left. The arrows indicate the gas velocity in a reference frame in which the central filament is roughly at rest. The photoheating during reionization strongly boosts the gas pressure. This also increases the absolute difference in pressure between dense structures and voids. After reionization these structures are over-pressurized and start to expand, which is visible in the velocity fields. The gas velocities near the edged of the filament point inward before reionization, but outward after reionization. The arrows are scaled such that a velocity of 10 km/s corresponds to a length of 20 ckpc/$h$.}
  \label{fig:pressure_vel}
\end{figure*}

\subsubsection{Spatially varying pressure smoothing of the IGM}
\label{sec:pressure_smooth}

The photoheating of the IGM during reionization strongly increases its temperature and gas pressure. The energy injected per baryon, and hence the temperature increase, depend on the spectrum of the ionizing radiation but are approximately independent of gas density. This results in a fairly flat temperature-density relation for most of the gas shortly after reionization (see Fig.~\ref{fig:rho-T}), with the exception of regions that have been strongly gravitationally heated by shocks and compression during structure formation. The spatial variation of the gas pressure is, hence, largely dominated by the density variation with dense regions having the highest pressure. Many of the smaller/lower density structures, for which the photoheating dominates over the gravitational heating, e.g., filaments and walls, will consequently be over-pressurized and will start to expand after their reionization. Post-processed radiative transfer simulations do not capture this effect as the coupling to the hydrodynamics is missing. With our hybrid scheme, we can instead study the pressure smoothing of the IGM during and after patchy reionization. The expansion of photoheated structures is illustrated in Fig.~\ref{fig:pressure_vel}, which shows the gas pressure and gas velocity field of a region in our $z_\textrm{r}=5.3$ patchy simulation before and after its reionization.

\begin{figure*}
  \centering
  \includegraphics[width=\linewidth]{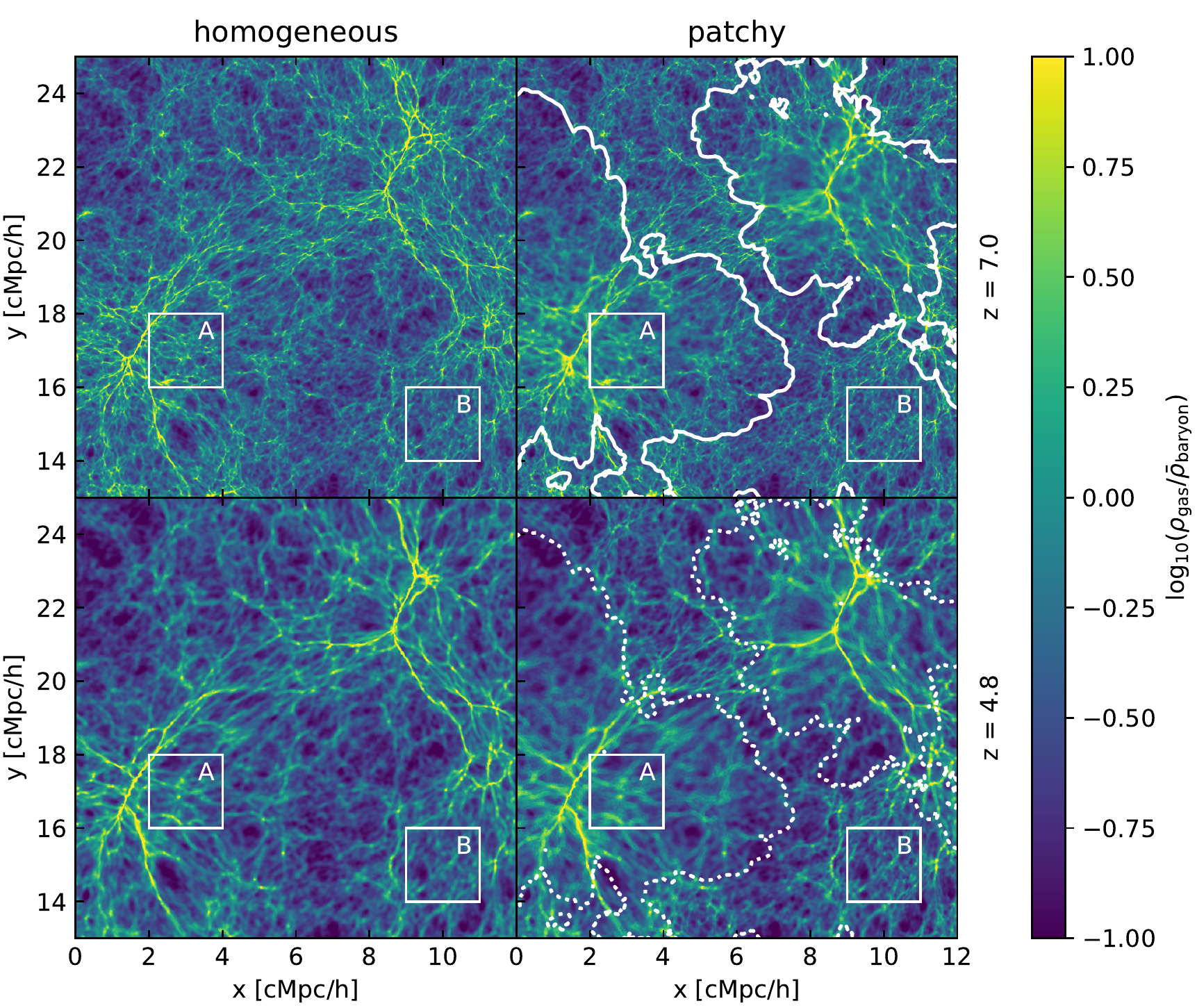}
  \caption{Gas density in units of the mean baryon density in a thin slice through our $z_\textrm{r}=5.3$ matched homogeneous and patchy simulations. The upper panels show results at redshift 7, when reionization is in full swing and both runs have a neutral fraction of $\sim 45\%$. In the upper right panel, regions that are already ionized are indicated by white contours and contain most large overdensities. Compared to the homogeneous simulation, the gas in the central regions of ionized bubbles has experienced more pressure smoothing in the patchy simulation. In contrast, regions that have not been ionized yet in the patchy simulation show more pronounced small scale structure than in the homogeneous run. This is illustrated in more detail for regions A and B for which zoom-ins are shown in Fig.~\ref{fig:slice_pressure_smooth_regions}. The bottom panels show the gas density in the same slices after reionization at redshift 4.8. The contours in the lower right panel indicate the same regions as in the upper right panel, hence separating regions that have reionized early (before redshift 7) from those that have reionized late. As illustrated by regions A and B, the difference in local reionization redshift results in notably different pressure smoothing even after reionization has ended (also see  Fig.~\ref{fig:slice_pressure_smooth_regions} for zoom-ins).}
  \label{fig:slice_pressure_smooth}
\end{figure*}

\begin{figure*}
  \centering
  \includegraphics[width=0.85\linewidth]{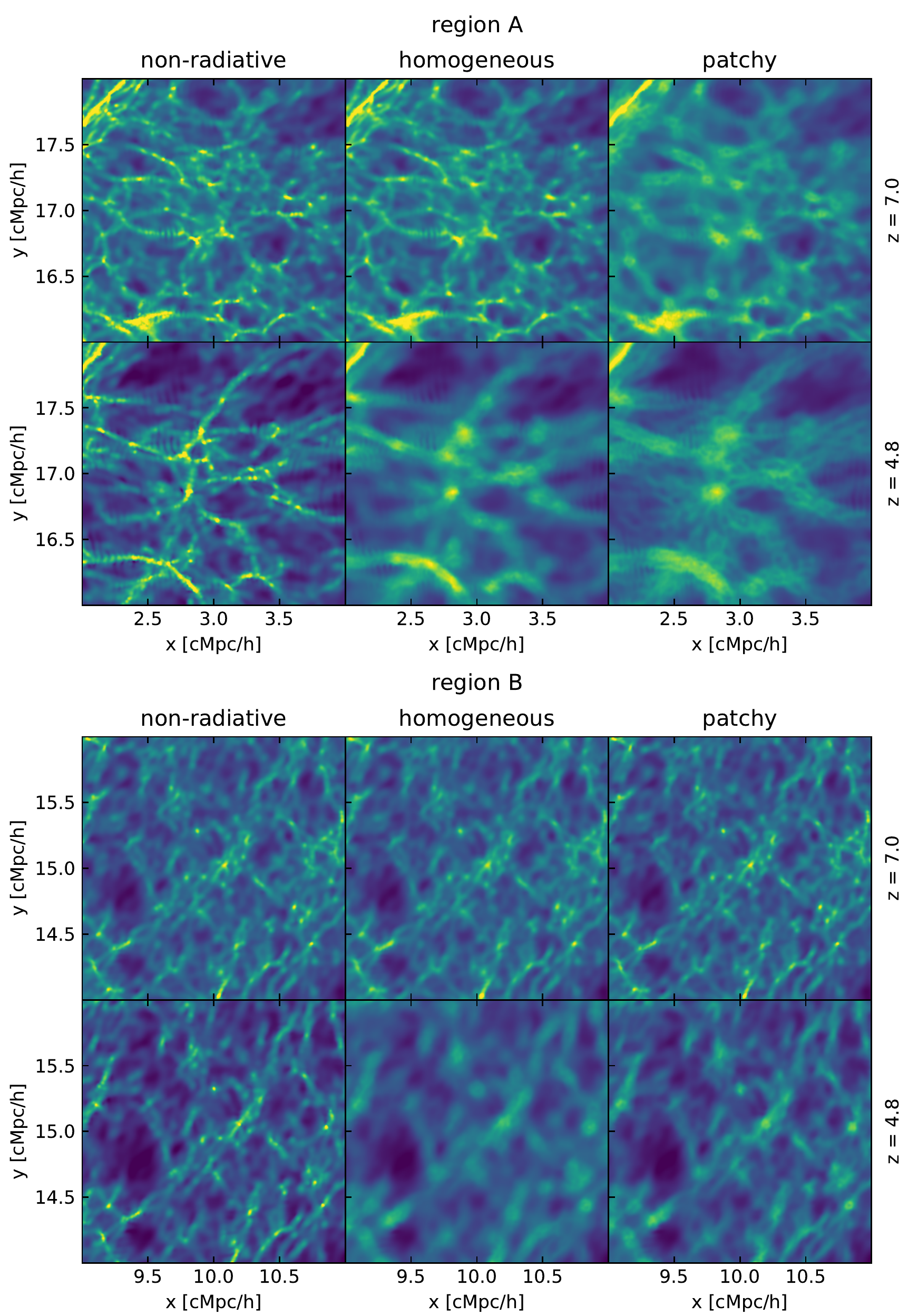}
  \caption{Gas density in units of the mean baryon density in thin slices covering regions that reionize early (region A, top set of panels) and late (region B, bottom set of panels) in our patchy reionization model. Density fields are shown at redshifts $7.0$ and $4.8$ for the $z_\textrm{r}=5.3$ patchy simulation (right panels), the matched homogeneous simulation (middle panels), and a non-radiative (``adiabatic'') simulation without photoheating and radiative cooling (left panels). The later is included to provide a reference model without pressure smoothing (due to photoheating). The color scale is the same as in Fig.~\ref{fig:slice_pressure_smooth}. The locations of regions A and B are also indicated there. Clearly early/late reionizing regions exhibit more/less pressure smoothing in the patchy simulation compared to the matched homogeneous model.}
  \label{fig:slice_pressure_smooth_regions}
\end{figure*}

These hydrodynamic reactions to the photoheating smooth the gas density distribution on small scales, roughly below the \textit{filtering scale} \citep[see][]{1998MNRAS.296...44G}, resulting in differences that persist well after reionization. Fig.~\ref{fig:slice_pressure_smooth} compares the gas density in a thin slice through our $z_\textrm{r}=5.3$ patchy simulation to that in the corresponding matched homogeneous simulation. This allows an assessment of how patchy reionization causes spatial fluctuations in the amount of small scale structure present in the gas density field.

The contours in the upper right panel, which shows the patchy run at $z=7$, indicate ionization fronts, i.e. the edges of ionized bubbles. Careful inspection shows that regions near the center of ionized bubbles (such as region A), which reionize early in the patchy simulation, are more strongly smoothed than in the matched homogeneous run where all regions largely follow the mean reionization history. Regions outside ionized bubbles (such as region B) have instead not experienced any pressure smoothing yet in the patchy run.

The bottom panels of Fig.~\ref{fig:slice_pressure_smooth} show the gas density in the same slice (in comoving coordinates) after reionization, at $z=4.8$. The dotted contours in the bottom right panel indicate the same regions as in the upper right panel, hence separating regions that have been reionized early (before $z=7$) from regions that have been reionized late (after $z=7$). Even at $z=4.8$, regions that have been reionized early have a smoother gas distribution than regions that have been reionized late.

To illustrate these effects more clearly, we zoom in on regions A and B in Fig.~\ref{fig:slice_pressure_smooth_regions}. For reference, we also show results for a non-radiative simulation of the same volume. This run does not include any photoionization or photoheating. Thus, no pressure smoothing is present outside shock-heated regions. This provides a reference model for comparison in which pressure smoothing is absent in the IGM.

Clearly, the early reionizing region A is most strongly smoothed in the patchy run at both redshifts. At $z=7$, the density field in the homogeneous run is still very similar to the non-radiative simulation, while pressure smoothing is clearly visible at $z=4.8$. In the patchy run at $z=4.8$, shells (visible as ring-like features) around photo-evaporated structures are visible. These will be discussed in more detail in Sec.~\ref{sec:lya_lines_rings}.

As expected, the neutral region B in the patchy simulation is indistinguishable from the non-radiative run at $z=7$. However, also the slice through the partly ionized homogeneous run still looks very similar. At $z=4.8$, region B is smoother in the homogeneous run compared to the patchy run. In the latter, the region reionizes late and has hence little time to respond to the heating.

In the following, we want to quantify the differences in the pressure smoothing that we have visually identified in the density fields. To this end, we perform local measurements of the power spectrum of the gas density contrast, $\delta = \Delta - 1 = \rho/\bar{\rho} -1$, where $\rho$ is the gas density and $\bar{\rho}$ the mean baryon density. For these measurements, we use a $2048^3$ grid covering the whole simulation box, and then randomly select 32768 regions of size $64^3$ grid cells, corresponding to a region side length of $1.25 \, \textrm{cMpc}/h$. The number of regions was chosen to sample a volume comparable to the full box. In each region, we then measure the gas density contrast power spectrum. We use a window function to reduce the impact of the non-periodic boundary conditions of the individual regions (see Appendix~\ref{app:pow_measure} for full details). We also compute the mean reionization redshift of each region, so that we can bin the power spectrum measurements by local reionization redshift. We perform this procedure both for the non-radiative simulation, as well as for our $z_\textrm{r}=5.3$ patchy run. We then compute for each reionization redshift bin the ratio of the mean power spectrum (averaged over all regions in the bin) in the patchy run to that (of the same regions) in the non-radiative (``adiabatic'') simulation. This quantifies the reduction of power caused by photoheating.

\begin{figure}
  \centering
  \includegraphics[width=\linewidth]{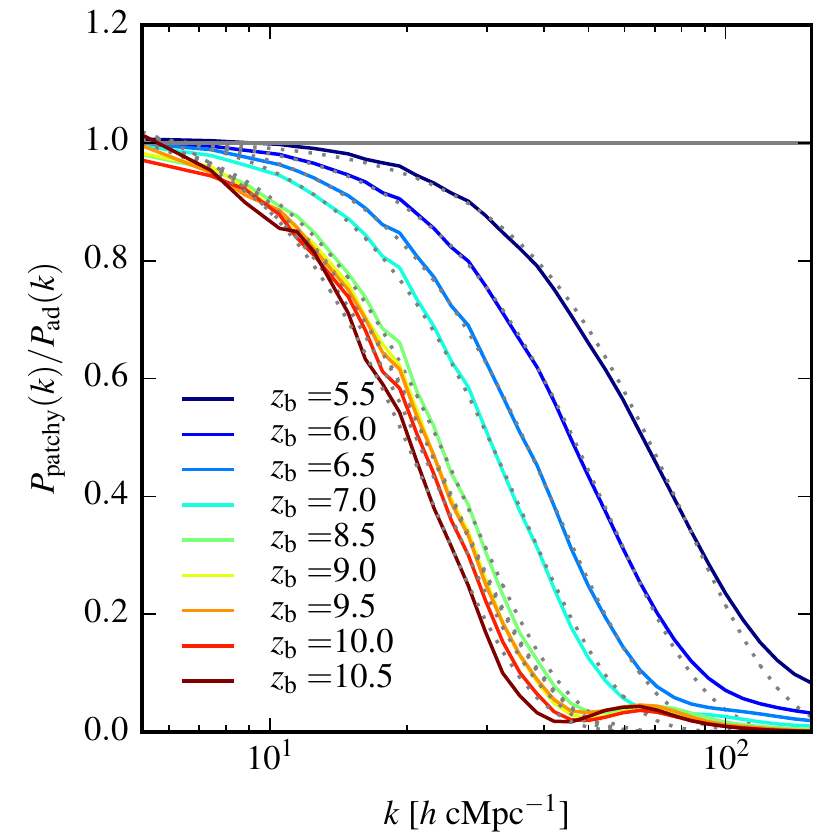}
  \caption{Impact of patchy reionization on small scale structure in the low-density IGM. Shown are ratios of the gas density ($\delta$) power spectra in the $z_\textrm{r}=5.3$ patchy and non-radiative (``adiabatic'') simulations. Results are shown at $z=4.8$ for regions with different mean reionization redshifts ($z_\mathrm{b}$ in the figure legend denotes the center of the $\Delta z=0.5$ bins into which the regions are sorted by their mean local reionization redshift). Regions that have been reionized earlier show a larger suppression of small scale power. All regions included here have a mean gas density of $0.2 < \Delta < 0.4$ (in units of the mean cosmic baryon density), corresponding to low-density IGM that the \lya\ forest is sensitive to. The dotted lines are fits to the solid curves assuming the functional form given in Eq.~(\ref{eq:cutoff}).}
  \label{fig:pow_spec_suppression}
\end{figure}

Fig.~\ref{fig:pow_spec_suppression} shows this quantity at $z=4.8$. Typically the gas density power spectrum is dominated by dense collapsed structures \citep[e.g.,][]{KulkarniEtAl2015}. As we are primarily interested in the low density IGM that is probed by the \lya\ forest at these redshifts, we opted to include only low-density regions with a mean density $0.2 < \Delta < 0.4$ in the computation of the mean power spectra. This is also a density range in which many of the shell/ring features discussed above (and further in Sec.~\ref{sec:lya_lines_rings}) reside and to which the \lya\ forest is sensitive to at very high redshifts. A reduction of gas density power compared to the non-radiative simulation is clearly present on small spatial scales (large $k$). We also find a clear dependence of the amount of suppression on the local reionization redshift of the considered regions. As expected, early reionizing regions show a suppression of power up to larger spatial scales (smaller $k$), while the regions that reionize latest (the $z_\textrm{r}=5.5$ bin) have the smallest reduction of small scale structure in the gas density field.

Overall the suppression of the power spectrum as a function of $k$ can be well described with a functional form similar to that used in \citet{1998MNRAS.296...44G}, i.e. with a suppression factor 
\begin{equation}
  \frac{P_\textrm{patchy}}{P_\textrm{ad}} = N \times \big(\exp(-k^2/k_\textrm{PS}^2)\big)^2,
\label{eq:cutoff}
\end{equation}
where $P_\textrm{patchy}$ and $P_\textrm{ad}$ are the power spectra in the patchy and non-radiative simulation respectively, and $k_\textrm{PS}$ describes a pressure-smoothing scale. We will fit this function to the curves in Fig.~\ref{fig:pow_spec_suppression} to extract the corresponding pressure smoothing scales. In these fits, we treat $k_\textrm{PS}$ as a free parameter. We will compare the scale measured from the simulation in this way to the \citet{1998MNRAS.296...44G} filtering scale (Eq.~\ref{eq:filtering_scale}) later in this section. To allow a bit more flexibility in the fits we have included a normalization factor $N$ which helps to absorb some of the effects that are caused by radiative cooling and star formation in dense objects in the patchy run. We expect this factor to be close to unity and indeed we find numerical values in the range 1.006 to 1.075 for the different reionization redshift bins. We perform the fits only in the range where $P_\textrm{patchy}/P_\textrm{ad} > 0.1$ to avoid being affected by the oscillations that are present at lower values (higher $k$) in the case of high reionization redshifts. We interpret these oscillations as an effect similar to that found in \citet{1998MNRAS.296...44G} for linear perturbations (see their fig. 1). We also note that while the functional form given by Eq.~(\ref{eq:cutoff}) works well for the suppression of power in low-density regions ($\Delta \sim 0.3$), it does so less well for regions at higher mean density ($\Delta \sim 1$ and larger), likely due to a larger number of non-linearly collapsed structures there.

\begin{figure}
  \centering
  \includegraphics[width=\linewidth]{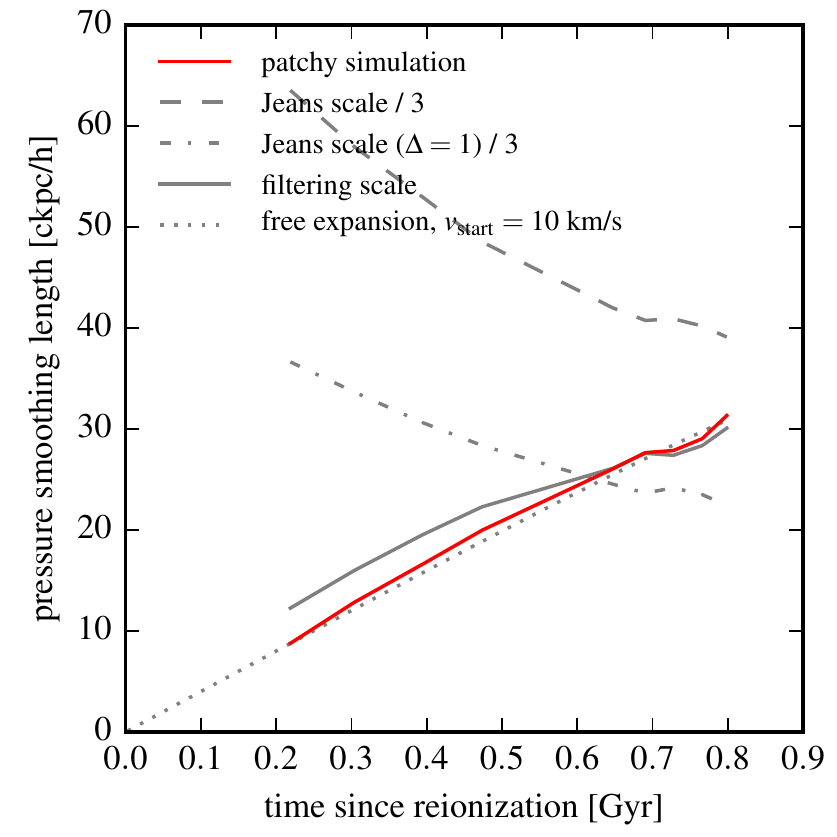}
  \caption{Pressure smoothing length scale at $\Delta \approx 0.3$ and $z=4.8$ as a function of time since reionization. The values for the patchy simulation were obtained from the parameters of the fits shown in Fig.~\ref{fig:pow_spec_suppression}.  For comparison we show the Jeans scale either evaluated at the actual density or at mean density ($\Delta = 1$). Note that the values of the Jeans scales were divided by 3 to better fit on the figure. Furthermore, we indicate the filtering scale computed as in \citet{1998MNRAS.296...44G}, as well as the length scale that corresponds to a free expansion with a starting velocity of 10 km/s (see main text for details).}
  \label{fig:lsmooth}
\end{figure}

The measured pressure smoothing scale $k_\textrm{PS}$ can then be converted to a length scale; we do this by defining the pressure smoothing length scale by $\lambda_\textrm{PS} \equiv 1/k_\textrm{PS}$. We do not include a $2 \pi$ factor in this definition to stay consistent with \citet{KulkarniEtAl2015}. The $\lambda_\textrm{PS}$ values inferred from the fits are shown by the red curve in Fig.~\ref{fig:lsmooth}. On the x-axis, we show the time since reionization, i.e. the time between the mean reionization redshift of the regions in a reionization redshift bin and the redshift, $z=4.8$, at which the power spectrum suppression is measured. As expected, the pressure smoothing scale increases with time since reionization as there is more time for the expansion of structures after their photoheating.

We next compare the measured pressure smoothing scale to different theoretical estimates such as the Jeans scale, the \citet{1998MNRAS.296...44G} filtering scale, and the distance corresponding to a simple free expansion with a fixed starting velocity.  The Jeans scale is an instantaneous measure which corresponds to the scale at which the sound crossing time matches the free fall time. Structures below this scale are typically assumed to be suppressed. Here, we define the co-moving Jeans scale similar to equation (2) of \citet{1998MNRAS.296...44G},
\begin{equation}
  k_\textrm{J} = \frac{a}{c_\textrm{s}} \sqrt{4 \pi G \bar{\rho}_\textrm{m} \Delta}
\end{equation}
but with an additional factor $\Delta$ to allow evaluation at different densities in units of the mean density. Here $c_\textrm{s} = \sqrt{(5 k T)/(3 \bar{m})}$ is the sound speed, with $T$ being the median temperature of a region, $k$ the Boltzmann constant and $\bar{m}$ the mean particle mass. $\bar{\rho}_\textrm{m}$ is the mean physical matter density at the considered redshift, and $a$ is the scale factor.

In linear perturbation theory, this should be evaluated at $\Delta=1$, but has a clear meaning only if the temperature evolves like $T\propto a^{-1}$ \citep[see][]{1998MNRAS.296...44G}. For different values of $\Delta$, this would correspond to matching sound crossing and free fall time at that density, but only in the absence of an expanding background. Here we are interested in low density regions, $\Delta \sim 0.3$, which often will expand even in the absence of a thermal pressure. It is thus not entirely clear how well motivated evaluating this at $\Delta \sim 0.3$ is. Nevertheless we show results for both, i.e. $\Delta$ set to the mean density of a region and $\Delta$ set to 1. We then average the Jeans length scale, $\lambda_\textrm{J} \equiv k_\textrm{J}^{-1}$, over all regions falling in a reionization redshift bin. Also, note that we have divided the Jeans length scales by a factor of 3 in Fig.~\ref{fig:lsmooth}, so that the curves fit better onto the plot.

Independent of the choice of $\Delta$, we find that the Jeans scale does not adequately reproduce the pressure smoothing scale measured from the simulation. This is not too surprising as an instantaneous measure cannot faithfully capture the time evolution of the pressure smoothing following a heating event. As we are measuring the pressure smoothing shortly after the end of reionization, when there was only limited time for the IGM to hydrodynamically react to the heating, we find that the Jeans scale is much larger than the measured pressure smoothing scale (keep in mind the division by 3). It also decreases, rather than increases, with the time since reionization as recently reionized regions are hot and have a correspondingly large Jeans scale.

Next, we compute the \citet{1998MNRAS.296...44G} filtering scale, which aims to capture the time evolution properly by taking the thermal history into account. The filtering scale $k_\textrm{F}$ is given by their equation (6), i.e. by
\begin{equation}
\frac{1}{k^2_\textrm{F}} = \frac{1}{D_+(t)} \int^t_0 dt' c^2_\textrm{s}(t') D_+(t') \int^t_{t'} \frac{dt''}{a^2(t'')}, \label{eq:filtering_scale}
\end{equation}
where $D_+(t)$ is the growth function of linear perturbations. As we are considering high redshifts here, $z \geq 4.8$, we simplify this by using $D_+ \propto a$ and approximating the Hubble function by $H \approx H_0 \sqrt{\Omega_\textrm{m} a^{-3}}$, where $H_0$ and $\Omega_\textrm{m}$ are the usual $\Lambda$CDM cosmological parameters. Using this, switching to $a' = a(t')$ as the integration variable and carrying out the inner integral, we can write the pressure smoothing scale as \citep[also see][]{2000ApJ...542..535G}
\begin{equation}
\frac{1}{k^2_\textrm{F}} = \frac{2}{a H_0^2 \Omega_\textrm{m}}
\int^a_0 da' a'^{\frac{3}{2}} \left[\frac{1}{\sqrt{a'}} - \frac{1}{\sqrt{a}}\right] \,
c^2_\textrm{s}(a').
\end{equation}
We evaluate this for each region using the history of the median gas temperature, convert this to a filtering length scale $\lambda_\textrm{F} \equiv k_\textrm{F}^{-1}$, and then average over all regions within the considered reionization redshift bin. The results of this are shown by the solid gray line in Fig.~\ref{fig:lsmooth}, which is overall in good agreement with the pressure smoothing scale measured from the simulation. This confirms that the \citet{1998MNRAS.296...44G} filtering scale describes the pressure smoothing of the low-density IGM well after reionization, and captures its time evolution.

Finally we compare the measured pressure smoothing scale to a simple expansion model in which structures expand freely after their photoionization/heating. This length scale in comoving units is given by
\begin{equation}
\lambda_\textrm{exp} = \int \frac{v_\textrm{start} a_\textrm{r}}{a(t)} \frac{dt}{a(t)} = v_\textrm{start} a_\textrm{r} \int_{a_\textrm{r}}^a \frac{da'}{a'^3 H(a')},
\end{equation}
where $v_\textrm{start}$ is the initial velocity right after reionization at $a_\textrm{r}$ and the $a_\textrm{r}/a$ term takes care of the cosmological decay of peculiar velocities. This expansion scale is shown for $v_\textrm{start} = 10 \, \textrm{km}/\textrm{s}$ in Fig.~\ref{fig:lsmooth}. Maybe somewhat surprisingly this simplistic model is in quite good agreement with the simulation. Note that the corresponding smoothing kernel for the gas density field, $\propto \exp(-k^2 \lambda_\textrm{exp}^2)$, corresponds to a Gaussian with standard deviation $\sigma = \sqrt{2} \lambda_\textrm{exp}$ in position space. Thus, the standard deviation of the real space smoothing kernel roughly grows like the travel distance for a $\sqrt{2} \times 10 \, \textrm{km}/\textrm{s} \approx 14 \, \textrm{km}/\textrm{s}$ starting velocity,  which is close to the speed of sound in a $\sim 10^4 \, \textrm{K}$ ionized IGM.

In particular, this simple free expansion model seems to reproduce the measured smoothing well and even better than the filtering scale shortly after reionization, $\lesssim 0.5 \, \textrm{Gyr}$. This suggests that the expansion of photoheated structures may not be strongly hindered by swept up material in the surrounding lower density regions in this time span. The lower level of agreement of the filtering scale may, however, be (partly) related to how the filtering scale is computed here. We use the whole thermal history of a region to compute the filtering scale. Part of a region will already be ionized and heated before the mean reionization redshift of the region is reached. Hence, despite using the median temperature, a region may already have a non-zero filtering scale at its mean reionization redshift, i.e. at a time of zero in Fig.~\ref{fig:lsmooth}. Such an offset could then still have a notable effect at somewhat later times. Furthermore, our measurement of the pressure smoothing in the simulation is based on a comparison of the patchy to the non-radiative run, thereby neglecting the small amount of pressure smoothing that is present in the latter. Also, we do not get the median temperature history directly from 3D grids but by using all pixels of a set of 5000 lines-of-sight through the simulation box that fall into a region. This gives us better time resolution ($\Delta z = 0.1$) as the line-of-sight files were saved more frequently than the full snapshots. It introduces, however, some noise which may contribute to such an offset. We have checked that using the line-of-sight file temperatures gives essentially the same result for the Jeans scale as the full temperature field, suggesting that the impact of this procedure on the filtering scale should also be small. Finally, we have neglected that the baryon density perturbation starts out at a much smaller value compared to the dark matter perturbation at the time of the decoupling of the CMB. This should, however, be a very small effect at the low redshifts considered here \citep{2022MNRAS.513..117L}.

\begin{figure*}
  \centering
  \includegraphics[width=0.9\linewidth]{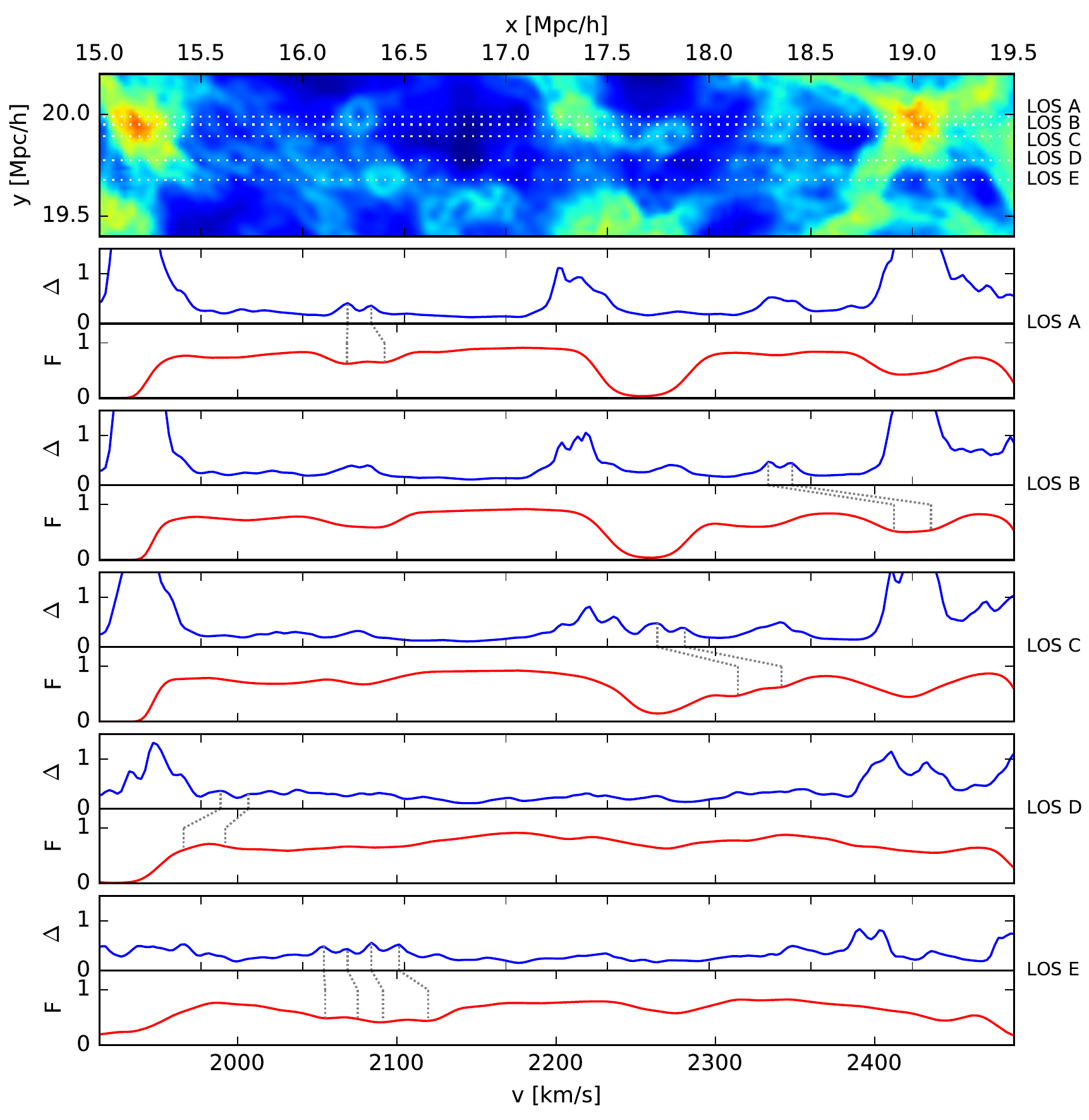}
  \caption{Expansion of gas overdensities due to the heating provided by reionization and their imprint on \lya\ forest absorption lines. The top panel displays the gas density in a thin slice in our $z_\textrm{r}=5.3$ patchy simulation at $z=4.2$. Also indicated are five lines-of-sight (LOS A-E) for which we explore the \lya\ transmission. The five other sets of panels show the density (in units of the mean) and the normalized transmitted flux for these five lines-of-sight. In the top panel various ``rings'' are visible. These appear when the slice cuts through expanding spherical or cylindrical shells that are produced when overdensities are evaporated by the photoheating provided by reionization. In the density skewers, these ``rings'' are visible as pairs of density peaks. Some of them are marked by gray dotted lines, which also connect to the corresponding redshift-space positions in the \lya\ spectra, i.e. the transmitted flux panels. There, isolated ``rings'' appear as absorbers with two minima (see, e.g., LOS A) or with a flat bottom (LOS B). ``Rings'' in regions with various other features in the density field can, e.g., appear as small dips in the larger scale features of the transmitted flux (e.g., LOS C-E). Such imprints can in principle be used to constrain pressure smoothing and reionization.}
  \label{fig:rings_specs}
\end{figure*}

Overall both the \citet{1998MNRAS.296...44G} filtering scale and the simple expansion model match the measured pressure smoothing in the underdense IGM well, while the Jeans scale is clearly inadequate shortly after reionization.

\subsubsection{\lya\ lines due to pressure-smoothed structures}
\label{sec:lya_lines_rings}

In Fig.~\ref{fig:slice_pressure_smooth_regions}, we have seen how pressure smoothing puffs up photoheated structures. This can result in shell-like features (visible as ``rings'' in thin 2D slices). Such features are also present in various other SPH and grid-based hydrodynamic simulations (see, e.g., fig.~9 in \citealt{KulkarniEtAl2015}, the highest resolution panel in fig.~8 of \citealt{2015MNRAS.446.3697L}, figs.~1 and 2 in \citealt{DAloisioEtAl2020}, fig.~4 in \citealt{2021ApJ...908...96P}, or fig.~6 in \citealt{2021ApJ...923..161N}), but have received only limited attention so far. Here, we investigate to what extent such shells and puffed up gas clouds leave a noticeable imprint in the \lya\ forest. The top panel of Fig.~\ref{fig:rings_specs} displays the gas density in a thin slice through part of our $z_\textrm{r} = 5.3$ patchy simulation at $z=4.2$. Various shell-like features, often visible as rings in this 2D slice, are present. To investigate their impact on \lya\ absorption, we shoot several lines-of-sight through the slice (dotted lines, labelled LOS A to E), and calculate synthetic \lya\ forest spectra for them. For each line of sight, we show a twin panel in Fig.~\ref{fig:rings_specs}, with the lower part showing the normalized transmitted \lya\ flux and the upper part showing the gas density in units of the mean baryon density. We have rescaled the optical depths by a constant factor to make the mean transmitted flux consistent with observed values (using Eq.~\ref{eq:tau_eff_fit} and computing the rescaling factor based on our full sample of 5000 lines-of-sight). The small dotted lines in the twin panels connect real-space positions of features in the density field to the corresponding redshift space position in the mock spectra. This facilitates identifying the associated absorption features.

LOS A passes through a ``ring'' in the density slice at $x\approx 16.3$ cMpc/$h$. A corresponding double peak is clearly visible in the density profile along the line-of-sight at that location (upper panel of the uppermost twin panel). The two peaks are marked by dotted lines. In redshift space, they correspond to a double-dip absorption  feature, which in this case directly reflects the shell traversed by the line-of-sight. Interestingly, the peak/dip separation is larger in redshift than in real space. This indicates that the shell is, as expected, expanding.

LOS B passes through a ``ring'' at $x\approx 18.3$ cMpc/$h$. Again, the double peak in the corresponding line-of-sight density profile is easily identified. In this case, the absorption features of the two peaks are more strongly overlapping, resulting in a broad flat bottom absorption profile in the mock spectrum. 

LOS C passes through a ``ring'' at $x\approx 17.8$ cMpc/$h$. While in the previous cases the double peak in the density profile was rather isolated, it falls near other structures here. This results in two absorption dips that fall on a larger scale gradient in the transmitted flux fraction. In an actual observation, this would likely make it more difficult to infer that these two dips originate from a single shell.

LOS D passes a less pronounced ``ring'' at $x\approx 15.7$ cMpc/$h$, and with several other shells and a larger structure nearby. Two corresponding small peaks can still be identified in the line-of-sight density profile, but the corresponding absorption features fall near a larger saturated absorber and are only visible as a slight change in the curvature of the spectrum. This would likely go unnoticed even in a high signal-to-noise observation.

LOS E passes through two ``rings'' at $x\approx 16.1$ and $16.4$ cMpc/$h$. Four corresponding peaks are visible in the line-of-sight density profile. Three of these cause visible dips in the synthetic spectrum. In an observation, it would, however, be difficult to identify which dips originate from the same shell, making inferring any pressure smoothing hard.

Overall, we find that shells caused by photoheated expanding structures often leave a direct imprint in the \lya\ forest, with distinct absorption dips at the shell ``walls''. In the most distinct case of a single, isolated, roughly spherical/cylindrical shell an intriguing double-dip absorption profile is imprinted on the \lya\ forest spectrum (see LOS A). Measuring, e.g., the separation of the two dips should in principle allow inferring the amount of pressure smoothing that the object has experienced. In practice, it may be difficult to identify which absorption features correspond to a shell originating from the same structure (see, e.g., LOS D and E), as well as what the exact orientation of the line-of-sight with respect to the shell is. An analysis to infer the pressure smoothing based on such features would hence likely require a larger sample of absorption systems along with a tailored statistical technique tested on simulations. It would also be important to check how well different hydrodynamics schemes agree on the prominence and properties of such shells. In addition, their abundance may depended on the amount of preheating of the neutral IGM by X-rays and the relative streaming velocity between baryons and dark matter \citep[see, e.g., fig.~4 of][]{2021ApJ...908...96P}. Finally, observations would need to be done at a suitable redshift at which the \lya\ forest is sensitive to the typical densities of such shells. The redshift considered here, $z=4.2$, seems to work reasonably well for this.   

\section{Summary and conclusions}
\label{sec:summary}

We have presented the Sherwood-Relics simulations, a new suite of cosmological hydrodynamical simulations aimed at modelling the IGM during and after cosmic reionization. The main difference to our previous Sherwood simulation project, which we build on in this work, is an improved treatment of the ionizing UV radiation field and of the thermo-chemistry of the IGM. 

Our new simulation sample consists of over 200 runs covering cubic volumes with sidelengths ranging from 5 to 160 cMpc/$h$. These are populated with between $2\times512^3$ and $2\times2048^3$ particles. Most of the simulations use an updated time-dependent but spatially homogeneous UV background model along with a non-equilibrium thermo-chemistry solver. These runs cover a wide range in thermal evolutions, cosmological parameters, dark matter free streaming scales and reionization histories, and will be instrumental for deriving constraints on these properties from \lya\ forest observations. 

The main focus of the analysis presented in this work is, however, the impact of a more realistic patchy cosmic reionization process on the properties of the IGM during the epoch of reionization, as well as its \textit{relic} signatures that persist for a considerable amount of time in the post-reionization IGM, such as spatial fluctuations in the IGM temperature, ionization state and small scale structure. To this end, we have developed a new hybrid radiative transfer/cosmological hydrodynamical simulation technique that allows following an inhomogeneous cosmic reionization process as well as the associated heating and pressure smoothing. The scheme uses radiation fields from post-processing radiative transfer simulations to photoionize and photoheat the IGM in a subsequent cosmological hydrodynamical simulation that then also captures the hydrodynamic response to the heating. This approach is suitable for the IGM,  computationally relatively cheap and circumvents the challenges of a full hydrodynamical modelling of the source galaxy population.

We assess the impact of the patchiness of reionization by comparing such ``patchy'' runs to homogeneous UVB simulations with the same mean reionization and thermal history. Our main findings are:
\begin{itemize}
\item Consistent with previous work, patchy reionization seeds IGM temperature fluctuations on large scales that persist well into the post-reionization epoch, down to $z\approx4$.
\item These temperature fluctuations are closely related to the local reionization redshift, with late reionizing regions being hotter.
\item The ionization state of the IGM reflects these temperature fluctuations. This causes a modulation of the \lya\ forest transmitted flux on large scales and a corresponding increase in the (one-dimensional) \lya\ forest power spectrum at $k\lesssim 10^{-2}$ s/km.
\item Patchy reionization also leads to a spatially varying pressure smoothing of the IGM. This results in spatial fluctuations in the amount of small scale density structure that is present in the IGM,  with early reionizing regions exhibiting the least amount of such structures.
\item Following reionization, the pressure smoothing length scale as a function of time since reionization is well described by the \citet{1998MNRAS.296...44G} filtering scale in the very low-density IGM. A simplistic free expansion model with an appropriate starting velocity also provides a reasonable fit. The instantaneous Jeans scale is instead not suitable for quantifying pressure smoothing shortly after reionization.
\item Pressure smoothing puffs up or evaporates small IGM structures such as filaments and small halos. This often results in shell-like features in the IGM density field and can leave characteristic imprints in the \lya\ forest, such as flat-bottom or double-dip absorption profiles.
\end{itemize}

These various impacts of the patchiness of cosmic reionization on the IGM and the high-redshift \lya\ forest should be taken into account when interpreting precision studies based on \lya\ forest data, in particular when using measurements at $z\gtrsim4$. A relatively simple way of doing this is to extract correction factors from patchy reionization simulations that can then be applied to (grids of) conventional homogeneous UVB simulations \citep[see, e.g.,][]{MolaroEtAl2022}. Alternatively, given the rather low computational cost of our hybrid patchy reionization simulation technique, tailored simulations can be performed to aid the interpretation of particular datasets.

The signatures of patchy reionization seen in our simulations may also be interesting for more accurately constraining the cosmic reionization process, e.g., by measuring the large scale increase of the \lya\ forest power spectrum as a function of redshift, or by analysing characteristic imprints of photoheated structures on the \lya\ forest. The latter would likely require the development of a suitable statistical technique to quantitatively compare such features between simulations and observations. The former would rely on more accurate measurements of the \lya\ forest power spectrum on large scales and near the epoch of reionization.

\section*{Acknowledgements}
We thank Volker Springel for making \textsc{p-gadget3} available and Debora Sijacki for helpful discussions. We acknowledge the Partnership for Advanced Computing in Europe (PRACE) for awarding us access to the Curie and Irene supercomputers, based in France at the Tr\`es Grand Centre de calcul du CEA, during the 16th Call.  The Sherwood-Relics simulations were also performed using time allocated during the Science and Technology Facilities Council (STFC) DiRAC 12th Call.  We used the Cambridge Service for Data Driven  Discovery (CSD3),  part of which is operated by the University of Cambridge Research Computing on behalf of the STFC DiRAC HPC Facility  (www.dirac.ac.uk).  The  DiRAC component of CSD3 was funded by BEIS capital funding via STFC capital grants ST/P002307/1 and ST/R002452/1 and STFC operations grant ST/R00689X/1.  This work also used the DiRAC@Durham facility managed by the Institute for Computational Cosmology on behalf of the STFC DiRAC HPC Facility. The equipment was funded by BEIS capital funding via STFC capital grants ST/P002293/1 and ST/R002371/1, Durham University and STFC operations grant ST/R000832/1. DiRAC is part of the National e-Infrastructure. JSB and MM are supported by STFC consolidated grant ST/T000171/1.  Laura Keating was supported by the European Union’s Horizon 2020 research and innovation programme under the Marie Skłodowska-Curie grant agreement No. 885990. GK is partly supported by the Department of Atomic Energy (Government of India) research project with Project Identification Number RTI 4002, and by the Max Planck Society through a Max Planck Partner Group. TŠ is supported by the University of Nottingham Vice Chancellor's Scholarship for Research Excellence (EU).  MV is supported by the INFN grant INDARK PD51 and by ASI grant under grant agreement ASI-INAF n. 2017-14-H.0. VI is supported by the Kavli Foundation. GDB is supported by the National Science Foundation through grant AST-1751404. Part of this work was supported by FP7 ERC Grant Emergence-320596. 

\section*{Data Availability}
The data and analysis code used in this work are available from the authors on request.  Further guidance for accessing  the publicly available Sherwood-Relics simulation data can be found on the project website:  \url{https://www.nottingham.ac.uk/astronomy/sherwood-relics/}




\bibliographystyle{mnras}
\bibliography{refs} 



\appendix

\section{Finding a homogeneous UVB model that yields the same reionization and thermal history as a patchy simulation}
\label{app:homogeneous_UVB}

To assess the impact of patchy reionization on various IGM properties and observables, it is useful to have a comparison run with a
homogeneous UV background that reproduces the ``mean'' ionization and thermal history of the patchy simulation. This isolates the impact of the ``patchiness'' of reionization.

As a first step we need to decide what kind of average of the inhomogeneously ionized and heated IGM we want to reproduce. Since we want to have comparable photoheating, we avoid high density regions in which shock heating plays a significant role. We therefore aim to match the IGM temperature at the mean cosmic baryon density in the patchy simulation. However, even gas at a fixed density can exhibit a wide range of temperatures during patchy reionization (see, e.g., Fig.~\ref{fig:rho-T}). Obvious choices of an ``average'' temperature are the mean or the median of the IGM temperature at mean density. The median has the advantage that it is less affected by shock heating of a small fraction of the gas to high temperatures. Unfortunately, during reionization, the median is almost a step function, increasing from very low temperatures to $\sim 8000$ K once the universe is $\sim50$ per cent ionized (see Fig.~\ref{fig:temp_xhii_taueff_vs_z}). Following such a sudden heating would be unreasonable in a simulation with a homogeneous UVB. To combine the advantages of both measures, we elect to follow the mean IGM temperature (at mean density) during reionization, but then switch to the median IGM temperature (at mean density) towards its end. As illustrated in Fig.~\ref{fig:temp_xhii_taueff_vs_z}, we switch at the time at which mean and median temperature are identical.

Furthermore, we want our comparison run with a homogeneous UVB to have a similar ionized fraction as the patchy simulation. We choose the mean ionized fraction of gas at mean cosmic baryon density in the patchy simulation as our target reionization history.

After measuring the target ionization and thermal history (as defined above) from the outputs of the patchy simulation (with a time resolution of $\Delta z = 0.1$), we apply some smoothing to them with a Savitzky-Golay filter to avoid following numerical noise in the evolution. The next step is then to compute photoionization and photoheating rates that reproduce the selected target ionization and thermal histories in a homogeneous simulation. To this end, we use a one-cell code that follows the thermal and ionization evolution of a single gas cell at mean cosmic baryon density. It is a modified version of the one-cell code described in appendix~C of \citet{PuchweinEtAl2019}. At each timestep, the code checks what hydrogen photoionization rate is necessary to continue following the target ionized hydrogen fraction. Similarly, it computes what photoheating rate is necessary to follow the target thermal history. Our \textsc{p-gadget3} version needs, however, not only the hydrogen rates as input, but also those for \hei\ and \heii. To get these we assume that the hydrogen and \hei\ photoionization rates match, i.e. $\Gamma_{\rm HeI}=\Gamma_{\rm HI}$. For the corresponding photoheating rates, we assume $\epsilon_{\rm HeI} = 1.3 \times \epsilon_{\rm HI}$ (roughly consistent with the time average of this ratio in the \citet{PuchweinEtAl2019} \textit{fiducial} UVB model). The \heii\ rates are simply adopted from the \citet{PuchweinEtAl2019} \textit{fiducial} UVB model. The latter have little impact during the hydrogen reionization epoch as significant \heii\ reionization happens only at lower redshift. The \hi, \hei, and \heii\ photoionization and photoheating rates obtained in this way as a function of redshift are then saved to a file, which can then be loaded into our \textsc{p-gadget3} version as a homogeneous UVB model. Simulations with this model then closely follow the chosen target ionization and thermal history (see Fig.~\ref{fig:temp_xhii_taueff_vs_z}).\\

\section{Local measurements of the gas density power spectrum}
\label{app:pow_measure}

In Sec.~\ref{sec:pressure_smooth}, we have performed local measurements of the power spectrum of the gas density contrast. To this end, we use the density contrast on a $2048^3$ grid covering the full simulation volume and then select 32768 regions of size $n_\textrm{reg}^3$ with $n_\textrm{reg} = 64$ from that grid for the local power spectrum measurement.

In contrast to the grid covering the full volume, the individual segments do not have periodic boundary conditions. To suppress the impact of this on the power spectrum measurement, we use a sine window function
\begin{equation}
w(l,m,n) = \sin\left(\frac{\pi \, l}{n_\textrm{reg}}\right)
\sin\left(\frac{\pi \, m}{n_\textrm{reg}}\right)
\sin\left(\frac{\pi \, n}{n_\textrm{reg}}\right)
\end{equation}
where $l$, $m$ and $n$ are the indices along the $x$, $y$ and $z$ direction of the cells in the segment that covers a region. They range from 0 to $n_\textrm{reg} - 1 = 63$. 
For correctly normalizing the power spectrum, we also need to compute the following sum \citep[see, e.g.,][]{PowSpecWindowFunction},
\begin{equation}
S_2 \equiv \sum_{l,m,n = 0}^{n_\textrm{reg}-1} w^2(l,m,n).
\end{equation}
We then calculate the power spectrum of a region by
\begin{equation}
P(|\vec{k}|) = \frac{\langle|\hat{\delta}_w|^2\rangle}{n_\textrm{reg}^3 S_2} L_\textrm{reg}^3,
\end{equation}
where $\langle|\hat{\delta}_w|^2\rangle$ is an average of $|\hat{\delta}_w|^2$ over all k-space points falling in the considered $|\vec{k}|$-bin used for the power spectrum computation. $\hat{\delta}_w$ is the discrete Fourier transform of $\delta_w = (\delta - \bar{\delta}_\textrm{r}) \times w$, where $\delta = \min(\rho_\textrm{IGM} / \bar{\rho}_\textrm{baryon} - 1, \,99)$ is the density contrast of the IGM normalized by the mean baryon density. We cap the density at a value of $1+\delta = \Delta \leq 100$ to reduce the impact of dense collapsed objects. $\bar{\delta}_\textrm{r}$ is the average of $\delta$ over the region. $L_\textrm{reg} = 1.25$ cMpc/$h$ is the sidelength of the region. The discrete Fourier transform is here defined without a normalization factor in the forward transform, i.e. by
\begin{equation}
\hat{f}(l',m',n') = \sum_{l,m,n = 0}^{n_\textrm{reg}-1} f(l,m,n) \, e^{\frac{i 2 \pi}{n_\textrm{reg}} (ll' + mm' + nn')},
\end{equation}
where $l'$, $m'$ and $n'$ are the indices of the k-space grid.

\begin{figure}
  \centering
  \includegraphics[width=\linewidth]{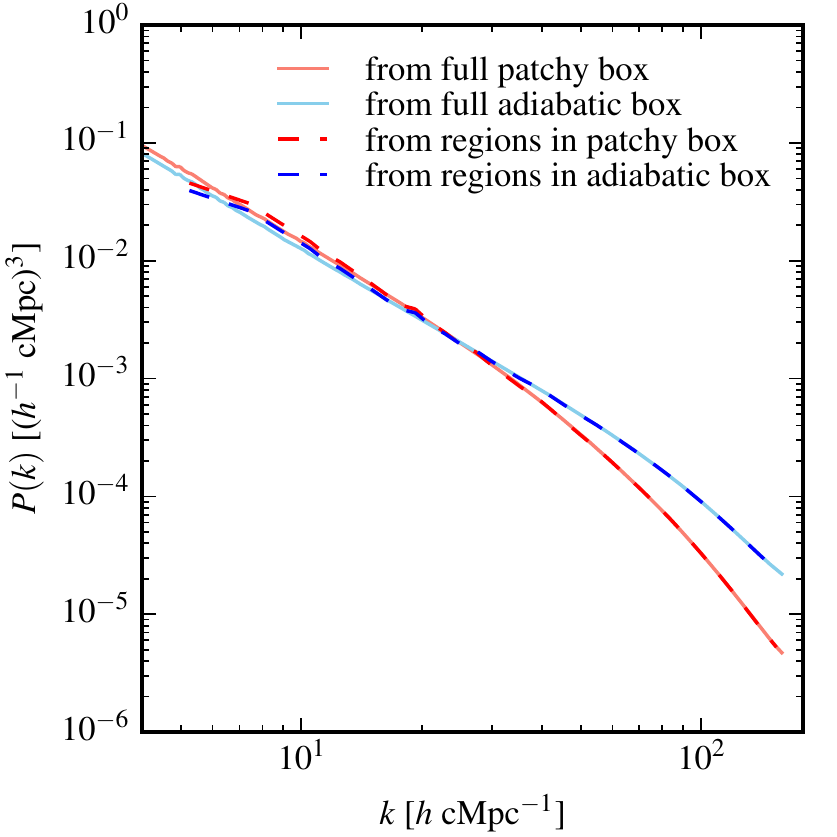}
  \caption{Gas density contrast power spectrum either calculated from a grid covering the full simulation box (solid lines) or by averaging the power spectra measured in all 32768 randomly placed regions (dashed lines). Results are shown for both the $z_\textrm{r}=5.3$ patchy and the adiabatic simulation at $z=4.8$. The averages of the local power spectrum measurements are in excellent agreement with the corresponding globally computed power spectrum.}
  \label{fig:pow_comp_reg_full}
\end{figure}

To test this procedure, we compute the average power spectrum of all 32768 randomly placed regions, i.e. without any cuts on mean density or local reionization redshift, and compare the results of this to the power spectrum calculated from the full grid covering the whole simulation volume. Fig.~\ref{fig:pow_comp_reg_full} displays this comparison for the $z_\textrm{r}=5.3$ patchy and the non-radiative/adiabatic simulation. We find good agreement on all overlapping $k$ scales. Small differences are visible for the smallest $k$ values (largest spatial scales) probed by the grids covering individual regions. This is expected as $k$-space is poorly sampled by only a handful of modes in the local power spectrum measurements there. On smaller spatial scales, where the pressure smoothing kicks in, the agreement is excellent. This confirms that the local power spectrum measurement works reliably.

\bsp	
\label{lastpage}
\end{document}